\documentclass[11pt, a4paper, oneside]{article}

\newcommand{\n}{\noindent}

\usepackage{amssymb}
\usepackage{amsfonts}

\begin{document}

\newpage
\setcounter{page}{0}
\begin{titlepage}
\begin{flushright}
UFSCARF-TH-06-11
\end{flushright}
\vskip 3.0cm
\begin{center}
{\Large  Reflection matrices for the $U_{q}[sl(r|2m)^{(2)}]$ vertex model }\\
\vskip 2cm
{\large A. Lima-Santos and W. Galleas } \\
\vskip 1cm
{\em Universidade Federal de S\~ao Carlos\\
Departamento de F\'{\i}sica \\
C.P. 676, 13565-905, S\~ao Carlos-SP, Brasil}\\
\end{center}
\vskip 2cm

\begin{abstract}
The graded reflection equation is investigated for the $U_{q}[sl(r|2m)^{(2)}]$ vertex model. We have found four classes of diagonal
solutions and twelve classes of non-diagonal ones. The number of free parameters for some solutions depends on the number
of bosonic and fermionic degrees of freedom considered.
\end{abstract}

\vskip 2.5cm
\centerline{{\small PACS numbers:  05.50+q, 02.30.IK, 75.10.Jm}}
\vskip 0.1cm
\centerline{{\small Keywords: Reflection Equations, K-matrices, Superalgebras}}
\vskip 2.5cm
\centerline{{\small June 2008}}
\end{titlepage}

\section{Introduction}

Two-dimensional integrable systems in statistical mechanics and quantum field theory have been a subject
of high importance in the last decades. In particular, systems of statistical mechanics with short range interactions
are believed to be conformally invariant at criticality \cite{POL} with significant implications in two-dimensions \cite{BEL,CAR}. More recently, a variety of integrable structures has emerged in the context of AdS/CFT correspondence \cite{ADS} in both sides of the duality. In the pioneer work \cite{MIN}, Minahan and Zarembo showed that the planar one-loop matrix of anomalous dimensions for a class of gauge invariant operators in the $\mathcal{N}=4$ Super Yang-Mills corresponds to the Hamiltonian of an integrable spin chain, which was latter generalized to all gauge invariant local operators in \cite{BEI}. Furthermore, in the string theory side of the correspondence, it was shown by Bena, Polchinksy and Roiban \cite{BPR} that the classical string theory sigma model on $AdS_{5} \times S^{5}$ is also integrable with evidences of integrability persisting at the quantum level \cite{JJ}.

Integrability in classical vertex models and quantum spin chains is intimately connected with solutions of the Yang-Baxter equation \cite{BAX}. This equation plays a central role in the Quantum Inverse Scattering Method which provides an unified approach to construct and study physical properties of integrable models \cite{QISM,KOR}. Usually these systems are studied with periodic boundary conditions but more general boundaries can also be included in this framework as well. Physical properties associated with the bulk of the system are not 
expected to be influenced by boundary conditions in the thermodynamical limit. Nevertheless, there are surface properties such as 
the interfacial tension where the boundary conditions are of relevance. Moreover, the conformal spectra of lattice models
at criticality can be modified by the effect of boundaries \cite{CAR1}.

Integrable models with open boundary conditions are also of interest in the context of AdS/CFT correspondence. Besides single trace 
operators, the gauge theory also contain baryonic operators, i.e. the so-called giant gravitons, corresponding to D-brane
excitations in the string counterpart \cite{WIT}. Such D-branes can appear in several circumstances \cite{CHEN,DEW,BER} and Berenstein and Vazquez have shown that the one-loop mixing of non-BPS giant gravitons can be described within the paradigm of integrable spin chains with
open boundary conditions \cite{BER}.

Integrable systems with open boundary conditions can also be accommodated within the framework of the Quantum
Inverse Scattering Method \cite{SK}. In addition to the solution of the Yang-Baxter equation governing the dynamics of the bulk
there is another fundamental ingredient, the reflection matrices \cite{CHER}. These matrices, also referred as $K$-matrices, represent the 
interactions at the boundaries and compatibility with the bulk integrability requires these matrices to satisfy the
so-called reflection equations \cite{SK,CHER}.

At the moment, the study of general regular solutions of the reflection equations for vertex models
based on $q$-deformed Lie algebras \cite{BAZ,JIM} has been successfully accomplished. See \cite{LIM} for instance 
and references therein. However, this same analysis for vertex models based on Lie superalgebras are still restricted
to diagonal solutions associated with the $U_{q}[sl(m|n)]$ \cite{G1,BRA} and $U_{q}[osp(2|2)]$ symmetries \cite{GUA} and non-diagonal solutions related to super-Yangians $osp(m|n)$ \cite{G2} and $sl(m|n)$ \cite{G3,GALLEAS5}.

The aim of this paper is to start to bridge this gap by presenting the most general set of solutions of the 
reflection equation for the $U_{q}[sl(r|2m)^{(2)}]$ vertex model. This paper is organized as follows. In the next section we
present the $R$-matrix of the $U_{q}[sl(r|2m)^{(2)}]$ vertex model in terms of standard Weyl matrices. This step paves the way
for the analysis of the corresponding reflection equations and in the section 3 we present what we hope to be the most general set
of $K$-matrices. Concluding remarks are discussed in the section 4, and in the appendices A and B we present special solutions associated
with the $U_{q}[sl(1|2)^{(2)}]$ and $U_{q}[sl(2|2)^{(2)}]$ cases respectively.

\section{The $U_{q}[sl(r|2m)^{(2)}]$ vertex model}

Classical vertex models of statistical mechanics are nowadays well known to play a fundamental role
in the theory of two-dimensional integrable systems \cite{BAX}. In this sense, it turns out that a 
$R$-matrix satisfying the Yang-Baxter equation gives rise to the Boltzmann weights of an
exactly solvable vertex model. The Yang-Baxter equation consist of an operator relation for a complex valued matrix
$R: \mathbb{C} \rightarrow \mbox{End}\left( V \otimes V \right)$ reading
\begin{equation}
\label{YB}
R_{12}(x) R_{13}(x y) R_{23}(y) = R_{23}(y) R_{13}(x y) R_{12}(y),
\end{equation}
where $R_{ij}(x)$ refers to the $R$-matrix acting non-trivially in the $i$th and $j$th spaces of
the tensor product $V \otimes V \otimes V$ and the complex variable $x$ denotes the spectral parameter.
Here $V$ is a finite dimensional $Z_{2}$ graded linear space and  the tensor products appearing in the above definitions should be understood in the graded sense. 
For instance, we have $\left[ A \otimes B  \right]^{\alpha \gamma}_{\beta \; \delta} = A^{\alpha}_{\beta} B^{\gamma}_{\delta} (-1)^{(p_{\alpha} + p_{\beta})p_{\gamma}}$ for generic
matrices $A$ and $B$. The Grassmann parities $p_{\alpha}$  assume values on the
group $Z_{2}$ and enable us to distinguish bosonic and fermionic degrees of freedom. 
More specifically, the $\alpha$th degree of freedom is distinguished by the
Grassmann parity
\begin{equation}
p_{\alpha}=\cases{
0  \;\;\;\;\; \mbox{for} \;\; \alpha \;\; \mbox{bosonic} \cr
1  \;\;\;\;\; \mbox{for} \;\; \alpha \;\; \mbox{fermionic} \cr }.
\end{equation}

An important class of solutions of the Yang-Baxter equation (\ref{YB}) is denominated 
trigonometric $R$-matrices containing an additional parameter $q$ besides the spectral parameter.
Usually such $R$-matrices have their roots in the $U_{q}[\mathcal{G}]$ quantum group framework, which
permit us to associate a fundamental trigonometric $R$-matrix to each Lie algebra or
Lie superalgebra  $\mathcal{G}$ \cite{BAZ,JIM,SHA}. In particular, explicit $R$-matrices were
exhibited in \cite{GALLEAS1,GALLEAS2} for a variety of quantum
superalgebras in terms of standard Weyl matrices, providing in this way a suitable basis for the analysis of the corresponding 
reflection equation.

\n The $U_{q}[sl(r|2m)^{(2)}]$ invariant $R$-matrices are given by
\begin{eqnarray}
\label{Rsl}
R(x) &=&\sum_{\stackrel{\alpha=1}{\alpha \neq \alpha'}}^{N} (-1)^{p_{\alpha}} a_{\alpha} (x)
\hat{e}_{\alpha \alpha} \otimes \hat{e}_{\alpha \alpha}
+b(x) \sum_{\stackrel{\alpha ,\beta=1}{\alpha \neq \beta,\alpha \neq \beta'}}^{N}
\hat{e}_{\alpha \alpha} \otimes \hat{e}_{\beta \beta}  \nonumber \\
&& +{\bar{c}} (x) \sum_{\stackrel{\alpha ,\beta=1}{\alpha < \beta,\alpha \neq \beta'}} (-1)^{p_{\alpha} p_{\beta}} \hat{e}_{\beta \alpha} \otimes  \hat{e}_{\alpha \beta}
+c(x) \sum_{\stackrel{\alpha ,\beta=1}{\alpha > \beta,\alpha \neq \beta'}}^{N} (-1)^{p_{\alpha} p_{\beta}} \hat{e}_{\beta \alpha} \otimes \hat{e}_{\alpha \beta} \nonumber \\
&& + \sum_{\alpha ,\beta =1}^{N} (-1)^{p_{\alpha}} d_{\alpha, \beta} (x)
\hat{e}_{\alpha \beta} \otimes \hat{e}_{\alpha' \beta'}
\end{eqnarray}
where $N=r+2m$ is the dimension of the graded space with $r$ bosonic and $2m$ fermionic degrees of freedom.
Here $\alpha' =N+1-\alpha$ corresponds to the conjugated index of $\alpha$ and $\hat{e}_{\alpha \beta}$
refers to a usual $N \times N$ Weyl matrix with only one non-null entry with value $1$ at the row $\alpha$ and column $\beta$. 

In \cite{GALLEAS3} it was demonstrated that the use of an appropriate grading structure plays a decisive role in the investigation of the
thermodynamic limit and finite size properties of integrable quantum spin chains based on superalgebras.
In what follows we shall adopt the grading structure
\begin{equation}
\label{grad}
p_{\alpha} =\cases{
\displaystyle  1 \;\;\;\; \mbox{for} \;\; \alpha=1,\dots ,m \;\;\; \mbox{and} \;\;\; \alpha =r+m+1,\dots ,r+2m \;\; \cr
\displaystyle  0 \;\;\;\; \mbox{for} \;\; \alpha=m+1,\dots ,r+m \;\; \cr },
\end{equation}
and the corresponding Boltzmann weights $a_{\alpha}(x)$, $b(x)$, $c(x)$, $\bar{c}(x)$ and $d_{\alpha \beta}(x)$ are then given by
\begin{eqnarray}
\label{bw1}
a_{\alpha} (x) &=&(x -\zeta)(x^{(1-p_{\alpha})} -q^2 x^{p_{\alpha}})  \;\;\;\;\;\;\;\;\;\;\;\;\;\;\;\; b (x) = q(x -1)(x -\zeta) \nonumber \\
c(\lambda) &=&(1-q^2)(x -\zeta)  \;\;\;\;\;\;\;\;\;\;\;\;\;\;\;\;\;\;\;\;\;\;\;\;\;\;\;\;\;\;\; {\bar{c}} (x) = x (1-q^2)(x -\zeta) \nonumber \\ 
\nonumber \\
d_{\alpha, \beta} (x)&=&\cases{
\displaystyle  q(x -1)(x -\zeta) +x(q^2 -1)(\zeta -1) \;\;\;\;\;\;\;\;\;\;\;\;\;\;\;\;\;\;\;\;\;\; \;\;\; \;\; \alpha=\beta=\beta' \;\; \cr
\displaystyle  (x -1)\left[ (x -\zeta) (-1)^{p_{\alpha}} q^{2 p_{\alpha}} +x(q^2 -1) \right] \;\;\;\; \;\;\;\; \;\;\; \;\;\;\;\;  \;\; \alpha=\beta \neq \beta' \;\; \cr
\displaystyle  (q^{2}-1)\left[ \zeta (x -1)\frac{\epsilon_{\alpha}}{\epsilon_{\beta}} q^{t_{\alpha}-t_{\beta}} -\delta_{\alpha ,\beta'} (x -\zeta) \right] \;\;\;\; \; \;\;\;\;\; \;\;\; \;\; \alpha < \beta \;\;\cr
\displaystyle  (q^{2 }-1) x \left[ (x -1)\frac{\epsilon_{\alpha}}{\epsilon_{\beta}} q^{t_{\alpha}-t_{\beta}} -\delta_{\alpha ,\beta'} (x -\zeta) \right] \;\;\; \;\;\;\;\;\;\; \;\;\;  \;\; \alpha > \beta \;\;\cr } 
\end{eqnarray}
where $\zeta=-q^{r-2m}$. The remaining variables $\epsilon_{\alpha}$ and $t_{\alpha}$
depend strongly on the grading structure considered and they are determined by the relations
\begin{eqnarray}
\epsilon_{\alpha} &=& \cases{
\displaystyle (-1)^{-\frac{p_{\alpha}}{2}} \;\;\;\;\;\;\;\; \;\;\;\; \;\;\;\;\;\;\;\;\;\;\; \;\;\;\;\;\;\;\;\;\;\;\;\; \;\; \;\; \;\; 1 \leq \alpha < \frac{N+1}{2} \;\; \cr
\displaystyle 1 \;\;\;\;\;\;\;\;\;\;\;\;\;\;\;\;\;\;\; \;\;\;\; \;\;\;\;\;\;\;\;\;\;\; \;\;\;\;\;\;\;\;\;\;\;\;\; \;\; \;\;  \;\; \alpha=\frac{N+1}{2} \;\; \cr
\displaystyle (-1)^{\frac{p_{\alpha}}{2}}  \;\;\;\;\;\;\; \;\;\;\; \;\;\;\;\;\;\;\;\;\;\;\;\;\;\;\;\;\;\;\;\;\;\; \;\;\;\;\;\;\;\;  \;\; \frac{N+1}{2} < \alpha \leq N \;\; \cr }
\\
t_{\alpha} &=& \cases{
\alpha + \left[ \frac{1}{2} -p_{\alpha} +2\displaystyle{\sum_{\alpha \leq \beta < \frac{N+1}{2}} p_{\beta}} \right] \;\;\;\;\;\; \;\;\;\;\;\; \;\;  \;\; 1 \leq \alpha < \frac{N+1}{2} \;\; \cr
\frac{N+1}{2} \;\;\; \;\;\;\;\;\;\;\;\;\;\;\;\;\;\;\;\;\;\;\;\;\;\;\;\;\;\;\;\;\;\;\;\;\;\;\;\;\;\;\;\;\;\;\;\;\;\;\;\;\;\;  \;\; \alpha = \frac{N+1}{2} \;\; \cr
\alpha - \left[ \frac{1}{2} -p_{\alpha} +2\displaystyle{\sum_{\frac{N+1}{2} < \beta \leq \alpha} p_{\beta}} \right] \;\;\;\;\;\;\;\;\;\;\;\; \;\;  \;\; \frac{N+1}{2} < \alpha \leq N \;\; \cr } .
\label{bwf}
\end{eqnarray}

The $R$-matrix (\ref{Rsl})-(\ref{bwf}) satisfies important symmetry relations, besides the standard properties of regularity and 
unitarity, namely
\begin{eqnarray}
&& \mbox{PT-Symmetry:} \;\;\;\;\;\;\;\;\;\;\;\;\;\;\;\; R_{21}(x)=R^{st_1 st_2}_{12}(x) \nonumber \\
&& \mbox{Crossing Symmetry:} \;\;\;\;\;\;\;\; R_{12}(x)=\frac{\rho(x)}{\rho(x^{-1} \eta ^{-1})}
V_{1} R_{12}^{st_2}(x^{-1} \eta ^{-1})V^{-1}_{1},  \nonumber 
\end{eqnarray}
where the symbol $st_k$ stands for the supertransposition operation in the space with index $k$.
In its turn $\rho(x)$ is an appropriate normalization function given by $\rho(x) =q(x -1)(x -\zeta)$
and the crossing parameter is $\eta = \zeta^{-1}$.
At this stage it is convenient to consider the $U_{q}[sl(2n|2m)^{(2)}]$ and  the $U_{q}[sl(2n+1|2m)^{(2)}]$ vertex models separately
and their corresponding crossing matrix $V$ is an anti-diagonal matrix with the following non-null entries $V_{\alpha {\alpha}^{'}}$,
\begin{itemize}
\item $U_{q}\left[ sl(2n|2m)^{(2)} \right]$:
\begin{eqnarray}
V_{\alpha {\alpha}^{'}}=\cases{
(-1)^{\frac{1-p_{\alpha}}{2}} q^{\left( \alpha -1+p_1-p_{\alpha} -2\displaystyle \sum_{\beta=1}^{\alpha-1} p_{\beta} \right) } \;\;\;\;\;\;\;\;\;\;\;\;\;\;\;\;\;\;\;  \;  1 \leq \alpha \leq \frac{N}{2} \cr
(-1)^{\frac{1+p_{\alpha}}{2}} q^{\left( \alpha -2 -p_1-p_{\alpha} -2\displaystyle \sum_{\stackrel{\beta=2}{\neq \frac{N}{2}+1}}^{\alpha-1} p_{\beta} \right) } \;\;\;\;\;\;\;\;\;\;\;\;\;\;\;\; \; \frac{N}{2}+1 \leq \alpha \leq N \cr}
\end{eqnarray}
\item  $U_{q}\left[sl(2n+1|2m)^{(2)} \right]$:
\begin{eqnarray}
V_{\alpha {\alpha}^{'}}=\cases{
(-1)^{\frac{1-p_{\alpha}}{2}} q^{\left( \alpha -1+p_1-p_{\alpha} -2\displaystyle \sum_{\beta=1}^{\alpha-1} p_{\beta} \right) } \;\;\;\;\;\;\;\;\;\;\;\;\;\;\;\;\;\;\;  \;  1 \leq \alpha \leq \frac{N-1}{2} \cr
(-1)^{\frac{1-p_{\alpha}}{2}} q^{ \left( \frac{N}{2} - 1 - p_1 - p_{\alpha} -2\displaystyle \sum_{\beta=2}^{\frac{N-1}{2}} p_{\beta} \right)} \;\;\;\;\;\;\;\;\;\;\;\;\;\;\;\;\;\;\;    \alpha=\frac{N+1}{2} \cr
(-1)^{\frac{1+p_{\alpha}}{2}} q^{\left( \alpha -2 -p_1-p_{\alpha} -2\displaystyle \sum_{\beta=2}^{\alpha-1} p_{\beta} \right) } \;\;\;\;\;\;\;\;\;\;\;\;\;\;\;\; \;\;\;\; \frac{N+3}{2}+1 \leq \alpha \leq N \cr}.
\end{eqnarray}
\end{itemize}

In the next section we shall investigate the possible regular solutions of the graded reflection equation for the $U_{q}[sl(r|2m)^{(2)}]$
vertex model.

\section{The $U_{q}[sl(r|2m)^{(2)}]$ reflection matrices}

The construction of integrable models with open boundaries was largely impulsed by Sklyanin's pioneer work \cite{SK}. In Sklyanin's 
approach the construction of such models are based on solutions of the so-called reflection equations \cite{CHER,SK} for a given
integrable bulk system.
The reflection equations determine the boundary conditions compatible with the bulk integrability and it reads
\begin{equation}
\label{RE}
R_{21}(x/y) K_{2}^{-}(x) R_{12}(xy) K_{1}^{-}(y) =
K_{1}^{-}(y) R_{21}(xy) K_{2}^{-}(x) R_{12}(x/y),
\end{equation}
where the tensor products appearing in (\ref{RE}) should be understood in the graded sense. The matrix $K^{-}(x)$ describes the 
reflection at one of the ends of an open chain while a similar equation should also hold for a matrix $K^{+}(x)$ describing the reflection at the opposite boundary. 
As discussed in the previous section, the $U_{q}[sl(r|2m)^{(2)}]$ $R$-matrix satisfies important symmetry relations such as the
PT-symmetry and crossing symmetry. When these properties are fulfilled one can follow the scheme devised in \cite{BRA,MEZ} and the 
matrix $K^{-}(x)$ is obtained by solving the Eq. (\ref{RE}) while the matrix $K^{+}(x)$ can be obtained from the isomorphism
$K^{-}(x) \mapsto K^{+}(x)^{st}=K^{-}(x^{-1} \eta^{-1}) V^{st} V$.

The purpose of this work is to investigate the general families of regular solutions of the graded reflection equation (\ref{RE}). Regular solutions mean that the $K$-matrices have the general form
\begin{equation}
\label{KM}
K^{-}(x)= \sum^{N}_{\alpha, \beta=1} k_{\alpha, \beta}(x) \; \hat{e}_{\alpha \beta},
\end{equation}
such that the condition $k_{\alpha ,\beta}(1)=\delta_{\alpha \beta}$ holds for all matrix elements.

The direct substitution of (\ref{KM}) and the $U_{q}[sl(r|2m)^{(2)}]$  $R$-matrix (\ref{Rsl})-(\ref{bwf})
in the graded reflection equation (\ref{RE}), leave us with a system of $N^{4}$ functional equations for the entries $k_{\alpha , \beta}(x)$. In order to solve these equations we shall make use of the derivative method. Thus, by differentiating the equation (\ref{RE}) with respect to $y$ and setting $y=1$, we obtain a set of algebraic equations for the matrix elements $k_{\alpha ,\beta}(x)$. Although we obtain a large number of equations only a few of them are actually independent and a direct inspection of those equations, in the lines described in \cite{LIM}, allows us to find the branches of regular solutions. In what follows we shall present our findings for the 
regular solutions of the reflection equation associated with the $U_{q}[sl(r|2m)^{(2)}]$ vertex model.
We have obtained four families of diagonal solutions and twelve families of non-diagonal ones. The special solutions associated with the
cases $U_{q}[sl(1|2)^{(2)}]$ and $U_{q}[sl(2|2)^{(2)}]$ are presented in the appendices A and B respectively.

\subsection{Diagonal solutions}
The diagonal solutions of the graded reflection equation (\ref{RE}) is characterized by a $K$-matrix of the form
\begin{equation}
K^{-}(x) = \sum_{\alpha=1}^{N} k_{\alpha, \alpha}(x) \hat{e}_{\alpha \alpha} .
\end{equation}
We have found four families of diagonal $K$-matrices for the $U_{q}[sl(r|2m)^{(2)}]$ vertex model that we shall refer as 
solutions of type $\mathcal{D}_{1}$, $\mathcal{D}_{2}$, $\mathcal{D}_{3}$ and $\mathcal{D}_{4}$. 
\begin{itemize}
\item {\bf Solution $\mathcal{D}_{1}$:} Solution with only one free parameter $\Omega$.
\end{itemize}
\begin{eqnarray}
k_{\alpha ,\alpha}(x) &=& \cases{
\displaystyle \frac{\Omega (x^{-1}-1)+2}{\Omega (x-1)+2} \;\;\;\;\;\;\;\;\;\;\;\;\;\;\;\;\;\;\;\;  \;\;  \alpha=1 \cr
\displaystyle 1 \;\;\;\;\;\;\;\;\;\;\;\;\;\;\;\;\;\;\;\;\;\;\;\;\;\;\;\;\;\;\;\;\;\;\;\;\;\;\;\;\;\;\;  \;\; \alpha=2, \dots, N-1 \cr
\displaystyle x\frac{\Omega (1-xq^{2}\zeta )-2}{\Omega (x-q^{2}\zeta )-2x} \;\;\;\;\;\;\;\;\;\;\;\;\;\;\;\;\;\;  \;  \alpha=N \cr}.
\end{eqnarray}

\vskip 1cm
\begin{itemize}
\item {\bf Solution $\mathcal{D}_{2}$:} Family formed by $m$ solutions without free parameters and
characterized by the label $p$ assuming discrete values in the interval $2 \leq p \leq m+1$.
\end{itemize}
\begin{eqnarray}
k_{\alpha ,\alpha}(x) &=& \cases{
\displaystyle  1  \;\;\;\;\;\;\;\;\;\;\;\;\;\;\;\;\;\;\;\;\;\;\;\;\;\;\;\;\;\;\;\;\;\;\;   \;\;  \alpha=1, \dots , p-1 \cr
\displaystyle \frac{x + \kappa q^{2p-3} \sqrt{\zeta}}{x^{-1} + \kappa q^{2p-3} \sqrt{\zeta}} \;\;\;\;\;\;\;\;\;\;\;\; \; \alpha=p, \dots , N+1-p \cr
\displaystyle x^2 \;\;\;\;\;\;\;\;\;\;\;\;\;\;\;\;\;\;\;\;\;\;\;\;\;\;\;\;\;\;\;\;\;  \;\; \alpha= N+2-p, \dots, N \cr}.
\end{eqnarray}
Here and in what follows, $\kappa$ is a discrete parameter assuming the values $\pm 1$.

\vskip 1cm
\begin{itemize}
\item {\bf Solution $\mathcal{D}_{3}$:} Solutions valid for $r \geq 4$ which does not contain free parameters. The discrete label $p$ can assume values in the interval $m+2 \leq p < \frac{N+1}{2}$.
\end{itemize}
\begin{eqnarray}
k_{\alpha ,\alpha}(x) &=& \cases{
\displaystyle  1  \;\;\;\;\;\;\;\;\;\;\;\;\;\;\;\;\;\;\;\;\;\;\;\;\;\;\;\;\;\;\;\;\;\;\;\;\;\;\;\;\;  \;\;  \alpha=1, \dots , p-1 \cr
\displaystyle \frac{x+\kappa q^{4m+1-2p}\sqrt{\zeta}}{x^{-1} + \kappa q^{4m+1-2p}\sqrt{\zeta}} \;\;\;\;\;\;\;\;\;\;\;  \;\; \alpha=p, \dots , N+1-p \cr
\displaystyle x^2 \;\;\;\;\;\;\;\;\;\;\;\;\;\;\;\;\;\;\;\;\;\;\;\;\;\;\;\;\;\;\;\;\;\;\;\;\;\;\;  \;\; \alpha= N+2-p, \dots, N \cr}.
\end{eqnarray}

\vskip 1cm
\begin{itemize}
\item {\bf Solution $\mathcal{D}_{4}$:} Class of solution valid only for $r=2n$  $(n \geq 1)$ with one free parameter $\Omega$.
\end{itemize}
\begin{eqnarray}
k_{\alpha ,\alpha}(x) &=& \cases{
\displaystyle  1  \;\;\;\;\;\;\;\;\;\;\;\;\;\;\;\;\;\;\;\;\;\;\;\;\;\;\;\;\;\;\;\;\;\;\;\;\;\;\;\;\;\;\;\;\;\;\;  \;\;  \alpha=1, \dots ,n+m-1 \cr
\displaystyle \frac{\Omega (x-1)+2}{\Omega (x^{-1} -1) +2}  \;\;\;\;\;\;\;\;\;\;\;\;\;\;\;\;\;\;\;\;\;\;\;\;\;\;\;  \alpha=n+m \cr
\displaystyle x \left[ \frac{ \Omega (x-q^2 \zeta^{-1}) - 2x}{\Omega (1-x q^2 \zeta^{-1}) - 2} \right] \;\;\;\;\;\;\;\;\;\;\;\;\; \;\;  \alpha=n+m+1 \cr
\displaystyle  x^2  \;\;\;\;\;\;\;\;\;\;\;\;\;\;\;\;\;\;\;\;\;\;\;\;\;\;\;\;\;\;\;\;\;\;\;\;\;\;\;\;\;\;\;\;\;  \;\;\;  \alpha=n+m+2, \dots ,N \cr}.
\end{eqnarray}

\subsection{Non-Diagonal Solutions}
Here we shall focus on the non-diagonal solutions of the graded reflection equation (\ref{RE}). We have found twelve classes of
non-diagonal solutions that we refer in what follows as solutions of type $\mathcal{N}_{1}$ to type $\mathcal{N}_{12}$.

\begin{itemize}
\item {\bf Solution $\mathcal{N}_{1}$:}
\end{itemize}
The solution of type $\mathcal{N}_{1}$ is valid only for the $U_{q}[sl(r|2)^{(2)}]$ model and the
$K$-matrix has the following block structure
\begin{equation}
K^{-}(x)=\pmatrix{
k_{1,1}(x) & \mathbb{O}_{1\times r} & k_{1,N}(x) \cr
\mathbb{O}_{r\times 1} & \mathbb{K}_{1}(x) & \mathbb{O}_{r\times 1} \cr
k_{N,1}(x) & \mathbb{O}_{1\times r} & k_{N,N}(x) \cr},
\end{equation}
where $\mathbb{O}_{a\times b}$ is a $a \times b$ null matrix and
\begin{equation}
\mathbb{K}_{1}(x) = \frac{x^2 - \zeta}{1-\zeta}  \; \mathbb{I}_{r \times r}.
\end{equation}
Here and in what follows $\mathbb{I}_{a \times a}$ denotes a $a \times a$ identity matrix and the remaining non-null entries are given by
\begin{eqnarray}
k_{1,1}(x)&=&1  \;\;\;\;\;\;\;\;\;\;\;\;\;\;\;\;\;\;\;\;\;\;\;\;\;\;\;\;\;\;\;\;\;\;\;\; k_{1,N}(x)=\frac{1}{2}\Omega (x^{2}-1) \nonumber \\
k_{N,1}(x)&=& \frac{2\zeta }{(1-\zeta )^{2}}\frac{(x^{2}-1)}{\Omega} \;\;\;\;\;\;\;\;\;\;\; k_{N,N}(x)=x^2,
\end{eqnarray}
where $\Omega$ is a free parameter.

\begin{itemize}
\item {\bf Solution $\mathcal{N}_{2}$:}
\end{itemize}
The $U_{q}[sl(r|4)^{(2)}]$ vertex model admits the solution $\mathcal{N}_{2}$ whose corresponding $K$-matrix
has the following structure
\begin{equation}
K^{-}(x)=\pmatrix{
\begin{array}{cc}
k_{1,1}(x) \quad & k_{1,2}(x) \quad \\ 
k_{2,1}(x) \quad & k_{2,2}(x) \quad
\end{array}
& \mathbb{O}_{2\times r} & 
\begin{array}{cc}
k_{1,N-1}(x) \quad & k_{1,N}(x) \quad \\ 
k_{2,N-1}(x) \quad & k_{2,N}(x) \quad 
\end{array} \cr 
\mathbb{O}_{r\times 2} & \mathbb{K}_{2}(x) & \mathbb{O}_{r\times 2} \cr
\begin{array}{cc}
k_{N-1,1}(x) & k_{N-1,2}(x) \\ 
k_{N,1}(x) & k_{N,2}(x)
\end{array}
& \mathbb{O}_{2\times r} & 
\begin{array}{cc}
k_{N-1,N-1}(x) & k_{N-1,N}(x) \\ 
k_{N,N-1}(x) & k_{N,N}(x)
\end{array}
\cr},
\end{equation}
where $\mathbb{K}_{2}(x) = k_{3,3}(x)  \mathbb{I}_{r \times r}$.
The non-diagonal entries can be written as
\begin{eqnarray}
k_{1,2}(x) &=&\Omega_{1,2}G_{1}(x)    \;\;\;\;\;\;\;\;\;\;\;\;\;\;\;\;\;\;\;\;\;\;\;\;\;\;\;\;\;\;\;\;  k_{2,1}(x)=\Omega_{2,1}G_{1}(x)  \nonumber \\
k_{1,N-1}(x) &=&\Omega_{1,N-1}G_{1}(x)   \;\;\;\;\;\;\;\;\;\;\;\;\;\;\;\;\;\;\;\;\;\;\;\;\;\;\;\;  k_{N-1,1}(x)=q^{2}\zeta
 \frac{\Omega_{1,2}\Omega_{2,1}}{\Omega_{1,N-1}}G_{1}(x)  \nonumber \\
k_{2,N-1}(x) &=&-\frac{\Omega_{2,1}\Omega_{1,N}}{\Omega_{1,2}} G_{1}(x)H(x)
\;\;\;\;\;\;\;\;\;\;\;\;\;\;
k_{N-1,2}(x)=-q^{2}\zeta \frac{\Omega_{1,2}\Omega_{2,1}\Omega_{1,N}}{\Omega_{1,N-1}^{2}}G_{1}(x)H(x)  \nonumber \\
k_{2,N}(x) &=&-\frac{\Gamma _{r}\Omega_{1,N}}{\Omega_{1,2}}xG_{2}(x)
\;\;\;\;\;\;\;\;\;\;\;\;\;\;\;\;\;\;\;\;\;\;\;
k_{N,2}(x)=-\frac{q^{2}\zeta \Gamma _{r}\Omega_{2,1}\Omega_{1,N}}{\Omega_{1,N-1}^{2}}xG_{2}(x) \nonumber \\
k_{N-1,N}(x) &=&-\frac{q^{2}\zeta \Gamma _{r}\Omega_{1,N}}{\Omega_{1,N-1}} xG_{2}(x)
\;\;\;\;\;\;\;\;\;\;\;\;\;\;\;\;\;\;
k_{N,N-1}(x)=-\frac{q^{2}\zeta \Gamma _{r}\Omega_{2,1}\Omega_{1,N}}{\Omega_{1,2}\Omega_{1,N-1}}xG_{2}(x) \nonumber \\
k_{N,1}(x) &=& q^{2}\zeta \frac{\Omega_{2,1}^{2}\Omega_{1,N}}{\Omega_{1,N-1}^{2} }G_{1}(x)H(x)
\;\;\;\;\;\;\;\;\;\;\;\;\;
k_{1,N}(x) = \Omega_{1,N} (x-1) ,
\end{eqnarray}
and the auxiliary functions are given by
\begin{eqnarray}
G_{1}(x)&=&\left[ -\frac{q^{2}\zeta \Gamma _{r}\Omega_{1,N}}{\Omega_{1,2}\Omega
_{1,N-1}}(x-1)+x-q^{2}\zeta \right] \frac{x-1}{x^{2}-q^{2}\zeta }, \nonumber \\
G_{2}(x)&=&\left[ -\frac{\Omega_{1,2}\Omega_{1,N-1}}{\Gamma _{r}\ \Omega_{1,N}}%
(x-1)+x-q^{2}\zeta \right] \frac{x-1}{x^{2}-q^{2}\zeta }, \\
H(x)&=&\frac{\Omega_{1,2}\Omega_{1,N-1}(x^{2}-q^{2}\zeta )}{\left[-q^{2}\zeta
\Gamma _{r}\Omega_{1,N}(x-1)+\Omega_{1,2}\Omega_{1,N-1}(x-q^{2}\zeta )\right]} \nonumber
\end{eqnarray}
and
\begin{equation}
\Gamma _{r}=\frac{\Omega_{2,1}\Omega_{1,N}}{\Omega_{1,N-1}}-\frac{2}{%
1-q^{2}\zeta }.
\end{equation}
With respect to the diagonal matrix elements, we have the following expressions
\begin{eqnarray}
k_{1,1}(x) &=&\left\{ \frac{q^{2}\zeta \Omega_{1,N}\Gamma _{r}^{2}}{\Omega_{1,2}\Omega_{1,N-1}}(1-q^{2}\zeta x)-(x-q^{2}\zeta )\left[ \frac{\Omega_{1,2}\Omega_{1,N-1}}{\Omega_{1,N}}-\frac{\Omega_{2,1}\Omega_{1,N}}{\Omega
_{1,N-1}}(1+q^{2}\zeta )\right] \right. \nonumber \\
&+&\left. \frac{2(1+q^{2}\zeta )^{2}x-4q^{2}\zeta (x^{2}+1)}{(1-q^{2}\zeta )(x-1)}%
\right\} \frac{(x-1)}{(x+1)(x^{2}-q^{2}\zeta )} \nonumber 
\end{eqnarray}
\begin{eqnarray}
k_{2,2}(x) &=& k_{1,1}(x)+ \Omega_{2,2} G_{1}(x) \nonumber \\
k_{3,3}(x) &=& k_{1,1}(x)+ \Omega_{3,3} G_{1}(x)+\Delta _{1}(x) \nonumber \\
k_{N-1,N-1}(x) &=& k_{3,3}(x) + \left(\Omega_{N-1,N-1}-\Omega_{3,3} \right) xG_{2}(x)+\Delta_{2}(x) \nonumber \\
k_{N,N}(x) &=& k_{N-1,N-1}(x) + \left(\Omega_{N,N}- \Omega_{N-1,N-1} \right)x G_{2}(x),
\label{dd}
\end{eqnarray}
where
\begin{eqnarray}
\Delta_{1}(x) &=&\frac{\Omega_{2,1}}{\Omega_{1,N-1}}\left[ x+\frac{q^{2}\zeta
\Gamma _{r}\Omega_{1,N}}{\Omega_{1,2}\Omega_{1,N-1}}\right] \frac{(x-1)^{2}}{
(x^{2}-q^{2}\zeta )}\Omega_{1,N} \nonumber \\
\Delta_{2}(x) &=&-\frac{q^{2}\zeta \Omega_{1,2}\Omega_{1,N-1}}{\Gamma_{r}\Omega_{1,N}} \Delta_{1}(x).
\end{eqnarray}
The diagonal entries (\ref{dd}) depend on the variables $\Omega_{\alpha , \alpha}$ which are related to the free parameters $\Omega_{1,2}, \Omega_{2,1}, \Omega_{1,N-1}$ and $\Omega_{1,N}$ through the expressions
\begin{eqnarray}
\Omega_{2,2} &=& \frac{\Omega_{1,2}\Omega_{1,N-1}}{\Omega_{1,N}}
-\Gamma_{r}  \;\;\;\;\;\;\;\;\;\;\;\;\;\;\;\;\; \Omega_{N-1,N-1} =2+\frac{\Omega_{1,2}\Omega_{1,N-1}}{\Omega_{1,N}}-q^{2}\zeta \Gamma_{r} \nonumber \\
\Omega_{3,3} &=& \frac{\Omega_{1,2}\Omega_{1,N-1}}{\Omega_{1,N}}+
\frac{2}{1-q^{2}\zeta} \;\;\;\;\;\;\;\;\;\;
\Omega_{N,N} =2+\frac{\Omega_{1,2}\Omega_{1,N-1}}{\Omega_{1,N}}-\frac{q^{2}\zeta \Gamma_{r}^{2}\Omega_{1,N}}{\Omega_{1,2}\Omega_{1,N-1}}. \nonumber \\
\end{eqnarray}

\vskip 1cm
\begin{itemize}
\item {\bf Solution $\mathcal{N}_{3}$:}
\end{itemize}
This class of solution is valid for the $U_{q}[sl(r|2m)^{(2)}]$ vertex model with $m \geq 3$ and the corresponding $K$-matrix possess the following general form 
\begin{equation}
K^{-}(x)=\pmatrix{
\begin{array}{ccc}
k_{1,1}(x)\quad & \cdots & k_{1,m}(x)\quad \\ 
\vdots \quad & \ddots & \vdots \quad \\ 
k_{m,1}(x)\quad & \cdots & k_{m,m}(x)\quad
\end{array}
& \mathbb{O}_{m\times r} & 
\begin{array}{ccc}
k_{1,r+m+1}(x)\quad & \cdots & k_{1,N}(x)\quad \\ 
\vdots \quad & \ddots & \vdots \quad \\ 
k_{m,r+m+1}(x) & \cdots & k_{m,N}(x)\quad
\end{array} \cr 
\mathbb{O}_{r\times m} & \mathbb{K}_{3}(x) & \mathbb{O}_{r\times m} \cr
\begin{array}{ccc}
k_{r+m+1,1}(x) & \cdots & k_{r+m+1,m}(x) \\ 
\vdots & \ddots & \vdots \\ 
k_{N,1}(x) & \cdots & k_{N,m}(x)%
\end{array}
& \mathbb{O}_{m\times r} & 
\begin{array}{ccc}
k_{r+m+1,r+m+1}(x) & \ldots & k_{r+m+1,N}(x) \\ 
\vdots & \ddots & \vdots \\ 
k_{N,r+m+1}(x) & \ldots & k_{N,N}(x)
\end{array}
\cr},
\end{equation}
where $\mathbb{K}_{3}(x)$ is a diagonal matrix given by
\begin{equation}
\mathbb{K}_{3}(x) = k_{m+1,m+1}(x) \; \mathbb{I}_{r \times r} .
\end{equation}

With respect to the elements of the last column, we have the following expression
\begin{eqnarray}
k_{\alpha,N}(x) &=&-\frac{\kappa }{\sqrt{\zeta }}\frac{\epsilon_{\alpha}}{\epsilon_{1}}
q^{t_{\alpha}-t_{1}}\Omega_{1,\alpha^{\prime }} x G(x)
\;\;\;\;\;\;\;\;\;\; \alpha=2, \dots ,m \;\; \mbox{and} \\
&& \;\;\;\;\;\;\;\;\;\;\;\;\;\;\;\;\;\;\;\;\;\;\;\;\;\;\;\;\;\;\;\;\;\;\;\;\;\;\;\;\;\;\;\;\;\;\;\;\;\;
\alpha=r+m+1, \dots ,N-1 , \nonumber
\end{eqnarray}
where $G(x)=x-1$. In their turn the entries of the first column are mainly given by
\begin{eqnarray}
k_{\alpha,1}(x)&=&\frac{\epsilon_{\alpha}}{\epsilon_{2}} q^{t_{\alpha}-t_{2}} \frac{\Omega_{2,1}\Omega_{1,\alpha^{\prime }}}{\Omega_{1,N-1}} G(x) \;\;\;\;\;\;\;\;\;\;\;\;\;\; \alpha=3, \dots ,m \;\; \mbox{and} \\
&& \;\;\;\;\;\;\;\;\;\;\;\;\;\;\;\;\;\;\;\;\;\;\;\;\;\;\;\;\;\;\;\;\;\;\;\;\;\;\;\;\;\;\;\;\;\;\;\;\;\;
\alpha=r+m+1, \dots ,N-1 . \nonumber
\end{eqnarray}
In the last row we have
\begin{eqnarray}
k_{N,\alpha}(x)&=&-\frac{\kappa }{\sqrt{\zeta }}\frac{\epsilon_{N} }{\epsilon_{2}}
q^{t_{N}-t_{2}}\frac{\Omega_{2,1}\Omega_{1,\alpha}}{\Omega_{1,N-1}} x G(x)
\;\;\;\;\;\;\;\;\;\;\;\; \alpha =2, \dots ,m \;\; \mbox{and} \\
&& \;\;\;\;\;\;\;\;\;\;\;\;\;\;\;\;\;\;\;\;\;\;\;\;\;\;\;\;\;\;\;\;\;\;\;\;\;\;\;\;\;\;\;\;\;\;\;\;\;\;\;\;\;\;\;\;\;\;\;
\alpha=r+m+1, \dots ,N-1 , \nonumber
\end{eqnarray}
while the elements of the first row are $k_{1, \alpha}(x) = \Omega_{1, \alpha} G(x)$ for $\alpha=2, \dots ,m$ and $\alpha=r+m+1, \dots ,N-1$. 

Concerning the elements of the secondary diagonal, they are given by
\begin{eqnarray}
k_{\alpha,\alpha^{\prime }}(x) &=& q^{t_{1}-t_{\alpha^{\prime }}}\frac{\epsilon_{1}}{\epsilon_{\alpha^{\prime }}}\frac{(1-\kappa q\sqrt{\zeta })^{2}}{(q+1)^{2}}\frac{\Omega_{1,\alpha^{\prime }}^{2}}{\Omega_{1,N}} G(x) H_{f}(x) \;\;\;\;\;\;\;\;\;\;\; 
\alpha=2, \dots, m \; , \; \alpha \neq \alpha' \nonumber \\
&& \;\;\;\;\;\;\;\;\;\;\;\;\;\;\;\;\;\;\;\;\;\;\;\;\;\;\;\;\;\;\;\;\;\;\;\;\;\;\;\;\;\;\;\;\;\;\;\;\;\;\;\;\;\;\;\;\;\;\;\;\;\;\;\;\;\;\;\;\;\;\;\; \alpha=r+m+1,\dots, N-1 ,
\end{eqnarray}
while the remaining entries $k_{1,N}(x)$ and $k_{N,1}(x)$ are determined by the following expressions 
\begin{eqnarray}
k_{1, N}(x) &=& \Omega_{1, N} G(x) H_{f}(x) \nonumber \\
k_{N, 1}(x) &=& \frac{\epsilon_{N-1}}{\epsilon_{2}}q^{t_{N-1}-t_{2}}\frac{\Omega_{1,N}\Omega_{2,1}^{2}}{\Omega_{1,N-1}^{2}} G(x) H_{f}(x) 
\end{eqnarray}
with $\displaystyle H_{f}(x) = \frac{x-\kappa q\sqrt{\zeta }}{1-\kappa q\sqrt{\zeta }}$.

The remaining matrix elements $k_{\alpha , \beta}(x)$ with $\alpha \neq \beta$ are then
\begin{equation}
k_{\alpha , \beta} (x) = \cases{
\displaystyle -\frac{\kappa}{\sqrt{\zeta}}\frac{\epsilon_{\alpha}}{\epsilon_{1}}
q^{t_{\alpha}-t_{1}}\left( \frac{1-\kappa q\sqrt{\zeta}}{q+1}\right) \frac{\Omega_{1,\alpha^{\prime }}\Omega_{1,\beta}}{\Omega_{1,N}} G(x)
\;\;\;\;\;\;\;\;\;\;\;\;\;\;\;\;\;\;\;\;\;\;  \alpha < \beta'  \;\;\; 2 \leq \alpha , \beta \leq N-1 \cr
\displaystyle \frac{1}{\zeta}\frac{\epsilon_{\alpha}}{\epsilon_{1}} q^{t_{\alpha}-t_{1}}
\left( \frac{1-\kappa q\sqrt{\zeta}}{q+1}\right) \frac{\Omega_{1,\alpha^{\prime
}}\Omega_{1,\beta}}{\Omega_{1,N}} x G(x)
\;\;\;\;\;\;\;\;\;\;\;\;\;\;\;\;\;\;\;\;\;\;\;\;\;\;  \alpha > \beta' \;\;\; 2 \leq \alpha , \beta \leq N-1 \cr
\displaystyle \frac{2(-1)^{m}q^{m-2}(1-\kappa q\sqrt{\zeta })}{(1+\kappa 
\sqrt{\zeta })(\kappa q^{m-1} \sqrt{\zeta }+(-1)^{m})(\kappa
q^{m} \sqrt{\zeta }+(-1)^{m})}\frac{\Omega_{1,N-1}}{\Omega_{1,N}} G(x)
\;\;\;\;\;\;\;\;\;\;\;\;\; \alpha=2, \; \beta=1 \cr
\displaystyle \frac{-2\zeta q^{m-1}(q+1)^{2}}{(1-\kappa q\sqrt{\zeta }
)(1+\kappa \sqrt{\zeta })(\kappa q^{m-1} \sqrt{\zeta }
+(-1)^{m})(\kappa q^{m} \sqrt{\zeta }+(-1)^{m})}\frac{\Omega_{1,N}}{\Omega_{1,N-m+1}} G(x) \cr
\;\;\;\;\;\;\;\;\;\;\;\;\;\;\;\;\;\;\;\;\;\;\;\;\;\;\;\;\;\;\;\;\;\;\;\;\;\;\;\;\;\;\;\;\;\;\;\;\;\;\;\;\;\;\;\;\;\;\;\;\;\;\;\;\;\;\;\;\;\;\;\;\;\;\;\;\;\;\;\;\;\;\;\;\;\;\;\;\;\;\;\;\;\;\;\;\;\;\;\;
\;\;\;\;\;\;\;\;\; \alpha=1, \; \beta=m \cr},
\end{equation}
and the parameters $\Omega_{1 , \alpha}$ are constrained by the relation
\begin{equation}
\Omega_{1,\alpha} = -\Omega_{1,\alpha+1}\frac{\Omega_{1,N-\alpha}}{\Omega_{1,N+1-\alpha}} 
\;\;\;\;\;\;\;\;\;\;\;\;\;\;\;\;\; \alpha=2, \dots, m-1 .
\end{equation}

With regard to the diagonal matrix elements, they are given by
\begin{equation}
k_{\alpha, \alpha}(x) = \cases{
\displaystyle \left( \frac{2x}{x^{2}-1} - \frac{\Omega_{m+1,m+1}}{x+1} \right) G(x)+\frac{\Delta_{1}(x)+\Delta_{2}(x)}{x^{2}-1} 
\;\;\;\;\;\;\;\;\;\;\;\;\;\;\;\;\;\;\;\;\;\;\;\;\;\;\; \alpha = 1 \cr
\displaystyle k_{1,1}(x)+ \Omega_{\alpha, \alpha} G(x) 
\;\;\;\;\;\;\;\;\;\;\;\;\; \;\;\;\;\;\;\;\;\;\;\;\;\;\;\;\;\;\; \;\;\;\;\;\;\;\;\;\;\;\;\;\;\;\;\;\;\;\;\;\;\;\;\;\;\;\;\;\;\;\;\;\;\;\;\;\;\;\; 2 \leq  \alpha \leq m  \cr
\displaystyle k_{1,1}(x)+ \Omega_{m+1,m+1} G(x)+\Delta _{1}(x) 
\;\;\;\;\;\;\;\;\;\;\;\;\;\;\;\;\;\;\;\;\;\;\;\;\;\;\;\;\;\;\;\;\;\;\;\;\;\;\;\;\;\;\;\;\;\;\;\;\; \alpha = m+1 \cr
\displaystyle k_{m+1,m+1}(x)+\left(\Omega_{r+m+1,r+m+1}-\Omega_{m+1,m+1} \right) x G(x)+\Delta _{2}(x)
\;\;\;\;\;\;\; \alpha = r+m+1 \cr
\displaystyle k_{\alpha-1,\alpha-1}(x) + \left( \Omega_{\alpha, \alpha}- \Omega_{\alpha-1, \alpha-1} \right)xG(x) 
\;\;\;\;\;\;\;\;\;\;\;\;\;\; \;\;\;\;\;\;\;\;\; \;\; r+m+2 \leq \alpha \leq N-1 \cr
\displaystyle x^{2}k_{1,1}(x) 
\;\;\;\;\;\;\;\;\;\;\;\;\;\;\;\;\;\;\;\;\;\;\;\;\;\;\;\;\;\;\;\;\;\;\;\;\;\;\;\;\;\;\;\;\;\;\;\;\;\;\;\;\;\;\;\;\;\;\;\;\;\;\;\;\;\;\;\;\;\;\;\;\; \alpha = N \cr}
\end{equation}
where
\begin{eqnarray}
\Delta_{1}(x) &=&\frac{2(x-1)^2}{(1+\kappa \sqrt{\zeta })(\kappa q^{m-1}
\sqrt{\zeta }+(-1)^{m})(\kappa q^{m}\sqrt{\zeta }+(-1)^{m})} ,  \nonumber \\
\Delta_{2}(x) &=& \kappa q^{r-1} \sqrt{\zeta }\Delta_{1}(x).
\end{eqnarray}
In their turn the diagonal parameters $\Omega_{\alpha ,\alpha}$ are fixed by the relations 
\begin{equation}
\Omega_{\alpha,\alpha} = \cases{
\displaystyle \Lambda_{m} \sum_{k=0}^{\alpha-2}(-\frac{1}{q})^{k}
\;\;\;\;\;\;\;\;\;\;\;\;\;\;\;\;\;\;\;\;\;\;\;\;\;\;\;\;\;\;\;\;\;\;\;\;\;\;\;\;\;\;\;\;\;\;\;\;\;\;\;\;\;\;\;\;\;\;\;\;\;\;\;\; 2 \leq \alpha \leq m  \cr
\displaystyle \Lambda_{m}\left[ \frac{q}{q+1}+\frac{(-1)^{m}
}{q^{m-2}}\frac{1}{(q+1)^{2}}\frac{1+\kappa \sqrt{\zeta }}{\kappa \sqrt{
\zeta }}\right] 
\;\;\;\;\;\;\;\;\;\;\;\;\;\;\;\;\;\;\;\;\;\;\;\;\; \alpha = m+1 \cr
\displaystyle 2 - \kappa q\sqrt{\zeta} \Lambda_{m} \sum_{k=0}^{N-\alpha-1}(-q)^{k}
\;\;\;\;\;\;\;\;\;\;\;\;\;\;\;\;\;\;\;\;\;\;\;\;\;\;\;\;\;\;\;\;\;\;\;\;\;\;\;\;\;\;\;\;\;\; r+m+1 \leq \alpha \leq N-1 \cr}
\end{equation}
with
\begin{equation}
\Lambda_{m}=\frac{2(-1)^{m}q^{m-2}(1+q)^{2}\kappa \sqrt{\zeta }}{(1+\kappa 
\sqrt{\zeta })(\kappa q^{m-1}\sqrt{\zeta }+(-1)^{m})(\kappa
q^{m} \sqrt{\zeta }+(-1)^{m})} .
\end{equation}
The class of solution $\mathcal{N}_{3}$ has a total amount of $m$ free parameters namely $\Omega_{1,r+m+1}, \dots, \Omega_{1,N}$.

\vskip 1cm
\begin{itemize}
\item {\bf Solution $\mathcal{N}_{4}$:}
\end{itemize}
This family of solutions is valid only for the $U_{q}[sl(2|2m)^{(2)}]$ vertex model and the corresponding $K$-matrix has the
following block structure
\begin{eqnarray}
K^{-}(x)=\pmatrix{
k_{1,1}(x) \mathbb{I}_{m \times m} & \mathbb{O}_{m \times 2} &  \mathbb{O}_{m \times m} \cr
\mathbb{O}_{2 \times m} & 
\begin{array}{cc}
k_{m+1,m+1}(x) & k_{m+1,m+2}(x) \\ 
k_{m+2,m+1}(x) & k_{m+2,m+2}(x)
\end{array}
& \mathbb{O}_{2 \times m} \cr
\mathbb{O}_{m \times m} & \mathbb{O}_{m \times 2} & k_{N,N}(x) \mathbb{I}_{m \times m} \cr}.
\end{eqnarray}
The non-null entries are given by
\begin{eqnarray}
k_{1,1}(x) &=& 1  \;\;\;\;\;\;\;\;\;\;\;\;\;\;\;\;\;\;\;\;\;\;\;\;\;\;\;\;\;\;\;\;\; k_{N,N}(x) = x^2  \nonumber \\
k_{m+1,m+1}(x) &=& \frac{x^{2}(1-\zeta )}{x^{2}-\zeta} \;\;\;\;\;\;\;\;\;\;\;\;\;\;\;\;\;\;\;\; k_{m+1,m+2}(x) =\frac{\Omega}{2}\frac{x(x^{2}-1)(1-\zeta )}{x^{2}-\zeta} \nonumber \\
k_{m+2,m+1}(x) &=& \frac{2}{\Omega}\frac{x(x^{2}-1)\zeta }{(1-\zeta
)(x^{2}-\zeta)} \;\;\;\;\;\;\;\; k_{m+2,m+2}(x) = \frac{x^{2}(1-\zeta )}{x^{2}-\zeta}
\end{eqnarray}
where $\Omega$ is a free parameter. We remark here that this solution for $m=1$ consist of a particular case of the 
solution given in the appendix B for the $U_{q}[sl(2|2)^{(2)}]$ vertex model.

\vskip 1cm
\begin{itemize}
\item {\bf Solution $\mathcal{N}_{5}$:}
\end{itemize}
The family $\mathcal{N}_{5}$ is acceptable by the vertex model $U_{q}[sl(3|2m)^{(2)}]$ and it is characterized by a 
$K$-matrix of the form
\begin{equation}
K^{-}(x)=\pmatrix{ 
k_{1,1}(x) \mathbb{I}_{m \times m} & \mathbb{O}_{m\times 3} & \mathbb{O}_{m\times m} \cr
\mathbb{O}_{3\times m} & 
\begin{array}{ccc}
k_{m+1,m+1}(x) & k_{m+1,m+2}(x) & k_{m+1,m+3}(x) \\ 
k_{m+2,m+1}(x) & k_{m+2,m+2}(x) & k_{m+2,m+3}(x) \\ 
k_{m+3,m+1}(x) & k_{m+3,m+2}(x) & k_{m+3,m+3}(x)
\end{array}
& \mathbb{O}_{3\times m} \cr 
\mathbb{O}_{m\times m} & \mathbb{O}_{m\times 3} & k_{N,N}(x) \mathbb{I}_{m \times m} \cr}.
\end{equation}
The non-diagonal matrix elements are given by the expressions
\begin{eqnarray}
k_{m+1,m+2}(x) &=& \Omega_{m+1,m+2} G(x) \;\;\;\;\;\;\;\;\;\;\;\;\;\;\;  k_{m+2,m+1}(x)=\Omega_{m+2,m+1} G(x) \nonumber \\
k_{m+3,m+2}(x) &=& \Omega_{m+3,m+2} xG(x) \;\;\;\;\;\;\;\;\;\;\;\;\; k_{m+2,m+3}(x)=\Omega_{m+2,m+3} xG(x)  \\
k_{m+1,m+3}(x) &=& \Omega_{m+1,m+3} G(x)H_{b}(x) \;\;\;\;\;\;\; k_{m+3,m+1}(x)=\Omega_{m+3,m+1}G(x)H_{b}(x) \nonumber 
\end{eqnarray}
where $H_{b}(x)=\frac{qx+\kappa \sqrt{\zeta }}{q+\kappa \sqrt{\zeta }}$ and $G(x)=x-1$. In their turn the above parameters $\Omega_{\alpha , \beta}$ are constrained by the relations
\begin{eqnarray}
\Omega_{m+2,m+3} &=&-i\kappa q^{m-1}\Omega_{m+1,m+2} \nonumber \\
\Omega_{m+3,m+2} &=&-i\kappa q^{m-1}\Omega_{m+2,m+1} \nonumber \\
\Omega_{m+3,m+1} &=&\Omega_{m+1,m+3} \left( \frac{\Omega_{m+2,m+1}}{\Omega_{m+1,m+2}} \right)^2
\end{eqnarray}
and
\begin{eqnarray}
\Omega_{m+2,m+1} &=& i\kappa q^{m-1}\frac{\Omega_{m+1,m+2}}{\Omega_{m+1,m+3}}
\left[ \frac{\sqrt{q}}{(q+1)}\frac{\Omega_{m+1,m+2}^{2}}{\Omega_{m+1,m+3}}-
\frac{2}{(q^{m-1/2}+i\kappa )(q^{m-3/2}-i\kappa )}\right]. \nonumber \\
\end{eqnarray}

The diagonal entries are then given by 
\begin{eqnarray}
k_{1,1}(x) &=&\left[ \frac{2}{x^{2}-1}+\frac{i\kappa }{q^{m-1}}\frac{
(xq^{2(m-1)}-1)}{x(x+1)}\Omega_{m+2,m+1}\frac{\Omega_{m+1,m+3}}{\Omega_{m+1,m+2}}\right] G(x)H_{b}(x) \nonumber \\
&+&i\kappa q^{m-1}\frac{\Omega_{m+1,m+2}^{2}}{\Omega_{m+1,m+3}}\frac{G(x)}{
x+1} \nonumber \\
k_{m+1,m+1}(x) &=&\left[ \frac{2}{x^{2}-1}+\frac{i\kappa }{q^{m-1}}\frac{
(q^{2(m-1)}+1)}{(x+1)}\Omega_{m+2,m+1}\frac{\Omega_{m+1,m+3}}{\Omega_{m+1,m+2}}\right] G(x)H_{b}(x) \nonumber \\
&+&i\kappa q^{m-1}\frac{\Omega_{m+1,m+2}^{2}}{\Omega_{m+1,m+3}}\frac{G(x)}{
x+1}
\end{eqnarray}
\begin{eqnarray}
k_{m+2,m+2}(x) &=&k_{m+1,m+1}(x)+\left( \Omega_{m+2,m+2}-\Omega_{m+1,m+1} \right) G(x)+\Delta (x) \nonumber \\
k_{m+3,m+3}(x) &=&k_{m+2,m+2}(x)+\left( \Omega_{m+3,m+3}-\Omega_{m+2,m+2} \right) xG(x) +\kappa \sqrt{\zeta}\Delta (x) \nonumber \\
k_{N,N}(x) &=& x^2 k_{1,1}(x) \nonumber
\end{eqnarray}
where
\begin{equation}
\Delta(x)=-\frac{q^{m-1}(x-1)^2}{(q^{m-1/2}+i\kappa )}
\frac{\Omega_{m+2,m+1} \Omega_{m+1,m+3}}{\Omega_{m+1,m+2}},
\end{equation}
and the parameters $\Omega_{m+1,m+1}, \Omega_{m+2,m+2}$ and $\Omega_{m+3,m+3}$ are fixed by the relations
\begin{eqnarray}
\Omega_{m+1,m+1} &=&\frac{i\kappa}{q^{m-1}}\Omega_{m+2,m+1}
\frac{\Omega_{m+1,m+3}}{\Omega_{m+1,m+2}} \nonumber \\
\Omega_{m+2,m+2} &=& \frac{2i\kappa }{q^{m-1/2}+i\kappa}
-i\kappa q^{m-1}\frac{\Omega_{m+1,m+2}^{2}}{\Omega_{m+1,m+3}} \\
\Omega_{m+3,m+3} &=&2-i\kappa q^{m-1}\Omega_{m+2,m+1}\frac{
\Omega_{m+1,m+3}}{\Omega_{m+1,m+2}}. \nonumber
\end{eqnarray}
The solution $\mathcal{N}_5$ possess two free parameters namely $\Omega_{m+1,m+2}$ and $\Omega_{m+1,m+3}$.

\vskip 0.5cm
\begin{itemize}
\item {\bf Solution $\mathcal{N}_{6}$:}
\end{itemize}
The solution $\mathcal{N}_{6}$ is admitted for the $U_{q}[sl(4|2m)^{(2)}]$ models with the following $K$-matrix
\begin{eqnarray}
\label{mm}
&& K^{-}(x)= \nonumber \\
&& \pmatrix{
k_{1,1}(x) \mathbb{I}_{m \times m} & \mathbb{O}_{m\times 4} & \mathbb{O}_{m\times m} \cr 
\mathbb{O}_{4\times m} & 
\begin{array}{cccc}
k_{m+1,m+1}(x) & k_{m+1,m+2}(x) & k_{m+1,m+3}(x) & k_{m+1,m+4}(x) \\ 
k_{m+2,m+1}(x) & k_{m+2,m+2}(x) & k_{m+2,m+3}(x) & k_{m+2,m+4}(x) \\ 
k_{m+3,m+1}(x) & k_{m+3,m+2}(x) & k_{m+3,m+3}(x) & k_{m+3,m+4}(x) \\ 
k_{m+4,m+1}(x) & k_{m+4,m+2}(x) & k_{m+4,m+3}(x) & k_{m+4,m+4}(x)
\end{array}
& \mathbb{O}_{4\times m} \cr
\mathbb{O}_{m\times m} & \mathbb{O}_{m\times 4} & k_{N,N} (x) \mathbb{I}_{m \times m} \cr} . \nonumber \\
\end{eqnarray}
The non-diagonal elements are all grouped in the $4 \times 4$ central block matrix. With respect to this central block, the entries of
the secondary diagonal are given by
\begin{eqnarray}
k_{m+1,m+4}(x) &=& (x-1) \Omega_{m+1,m+4} \nonumber \\
k_{m+2,m+3}(x) &=&-\frac{\Omega_{m+2,m+1}}{\Omega_{m+1,m+2}} \Omega_{m+1,m+4} (x-1) \nonumber  \\
k_{m+3,m+2}(x) &=&-\frac{q^{2}}{\zeta }\frac{\Omega_{m+1,m+2}\Gamma _{m}^{2}}{\Omega_{m+2,m+1}\Omega_{m+1,m+4}}(x-1) \\
k_{m+4,m+1}(x) &=&\frac{q^{2}}{\zeta }\frac{\Gamma _{m}^{2}}{\Omega_{m+1,m+4}}(x-1), \nonumber
\end{eqnarray}
and the remaining non-diagonal elements can be written as
\begin{eqnarray}
k_{m+1,m+2}(x)&=&\Omega_{m+1,m+2}G_{1}(x) \;\;\;\;\;\;\;\;\;\;\;\;\;\;\;\;\;\;\;\;\;\;\;\;\; k_{m+2,m+1}(x)=\Omega_{m+2,m+1}G_{1}(x) \nonumber \\
k_{m+1,m+3}(x)&=&\Omega_{m+1,m+3}G_{2}(x) \;\;\;\;\;\;\;\;\;\;\;\;\;\;\;\;\;\;\;\;\;\;\;\;\; k_{m+3,m+1}(x)=\frac{q^{2}}{\zeta}\frac{\Omega_{m+1,m+2}\Omega_{m+1,m+3}\Gamma_{m}^{2}}{\Omega_{m+2,m+1}\Omega_{m+1,m+4}^{2}}G_{2}(x) \nonumber \\
k_{m+2,m+4}(x) &=&-\frac{\Omega_{m+2,m+1}\Omega_{m+1,m+4}}{\Gamma_{m}}xG_{1}(x) \;\;\;\;\;\;\;\;\;\;\;\;\;\;\;
k_{m+4,m+2}(x) =-\frac{q^{2}}{\zeta}\frac{\Omega_{m+1,m+2}\Gamma_{m}}{\Omega_{m+1,m+4}} xG_{1}(x) \nonumber \\
k_{m+3,m+4}(x)&=&-\frac{q^{2}}{\zeta}\frac{\Omega_{m+1,m+2}\Omega_{m+1,m+3} \Gamma_{m}}{\Omega_{m+1,m+4}\Omega_{m+2,m+1}}xG_{2}(x) \;\;\;\;\;\;\;
k_{m+4,m+3}(x)=-\frac{q^{2}}{\zeta}\frac{\Omega_{m+1,m+3} \Gamma_{m}}{\Omega_{m+1,m+4}} xG_{2}(x), \nonumber \\
\end{eqnarray}
where 
\begin{eqnarray}
G_{1}(x)&=&\left[ \frac{\zeta -q^{2}x}{q^{2}(x-1)}+\frac{\Omega_{m+1,m+3}\Gamma
_{m}}{\Omega_{m+2,m+1}\Omega_{m+1,m+4}}\right] \frac{q^{2}(x-1)^2}{\zeta
-q^{2}x^{2}} , \nonumber \\
G_{2}(x)&=&\left[ \frac{\zeta -q^{2}x}{x-1}+\frac{\zeta \Omega_{m+2,m+1}\Omega_{m+1,m+4}}{\Omega_{m+1,m+3}\Gamma _{m}}\right] \frac{(x-1)^2}{\zeta -q^{2}x^{2}}
\end{eqnarray}
and
\begin{equation}
\Gamma_{m}=\frac{\Omega_{m+1,m+2}\Omega_{m+1,m+3}}{\Omega_{m+1,m+4}}+\frac{2\zeta}{q^{2}-\zeta}.
\end{equation}
In their turn the diagonal entries are given by the following expressions
\begin{eqnarray}
k_{1,1}(x) &=&\left\{ \frac{\zeta -q^{2}x}{(x+1)(\zeta -q^{2}x^{2})}\left[ \frac{
\Omega_{m+1,m+2}\Omega_{m+2,m+1}}{\Gamma_{m}} \right. \right. \nonumber \\
&& + \left. \left. \frac{\Omega_{m+1,m+2}\Omega_{m+1,m+3}}{\zeta \Omega_{m+1,m+4}}
\left( \frac{q^{2}\Omega_{m+1,m+3}\Gamma _{m}}{\Omega_{m+1,m+4} \Omega_{m+2,m+1}}+\frac{(\zeta +q^{2}x^{2})}{x}\right) \right] \right. \nonumber \\
&&+\left. \frac{2\left( \zeta -q^{2}x^{2}\right) }{x\left( x^{2}-1\right)
\left( \zeta -q^{2}\right) }\right\} (x-1) \nonumber
\end{eqnarray}
\begin{eqnarray}
k_{m+1,m+1}(x) &=& k_{1,1}(x)-\Gamma_{m} G_{1}(x)+\Delta_{1}(x) \nonumber \\
k_{m+2,m+2}(x) &=&k_{m+1,m+1}(x)+(\Omega_{m+2,m+2}+\Gamma_{m})G_{1}(x)  \\
k_{m+3,m+3}(x) &=&k_{m+2,m+2}(x)+\Delta_{2}(x)  \nonumber \\
k_{m+4,m+4}(x) &=&k_{m+3,m+3}(x)+(\Omega_{m+4,m+4}-\Omega_{m+3,m+3})xG_{2}(x) \nonumber \\
k_{N, N}(x) &=& x^2 k_{1,1} (x) \nonumber
\end{eqnarray}
where
\begin{eqnarray}
\Delta _{1}(x)&=&\frac{\Gamma _{m}\left( \Gamma _{m}\Omega_{m+1,m+3}q^{2}x+\zeta
\Omega_{m+1,m+4}\Omega_{m+2,m+1}\right) }{\Omega_{m+1,m+4}\Omega_{m+2,m+1}}\frac{(x-1)^2}{x(\zeta -q^{2}x^{2})} \nonumber \\
\Delta _{2}(x)&=&\frac{\Omega_{m+1,m+2}\left( \zeta \Omega_{m+2,m+1}^{2}\Omega_{m+1,m+4}^{2}-q^{2}\Omega_{m+1,m+3}^{2}\Gamma _{m}^{2}\right) }{\Gamma
_{m}\Omega_{m+1,m+4}^{2}\Omega_{m+2,m+1}}\frac{x(\zeta -q^{2})(x-1)}{
\zeta (\zeta -q^{2}x^{2})} . \nonumber \\
\end{eqnarray}
The variables $\Omega_{\alpha ,\alpha}$ are given in terms of the free parameters $\Omega_{m+1,m+2}, \Omega_{m+2,m+1}, \Omega_{m+1,m+3}$
and $\Omega_{m+1,m+4}$ through the relations
\begin{eqnarray}
\Omega_{m+2,m+2}&=&\frac{2\zeta }{\zeta -q^{2}}-\frac{\Omega_{m+1,m+2}\Omega_{m+2,m+1}}{\Gamma _{m}} \nonumber \\
\Omega_{m+3,m+3}&=&\frac{2\zeta }{\zeta -q^{2}}-\frac{q^{2}}{\zeta }\frac{\Gamma_{m}\Omega_{m+1,m+2}\Omega_{m+1,m+3}^{2}}{\Omega_{m+2,m+1}\Omega_{m+1,m+4}^{2}} \\
\Omega_{m+4,m+4} &=& \frac{2\zeta }{\zeta -q^{2}}-\frac{q^{2}}{\zeta }\frac{\Omega_{m+1,m+2}\Omega_{m+1,m+3}}{\Omega_{m+1,m+4}} . \nonumber
\end{eqnarray}

\vskip 1cm
\begin{itemize}
\item {\bf Solution $\mathcal{N}_{7}$:}
\end{itemize}

The vertex model $U_{q}[sl(2n|2m)^{(2)}]$ admits the solution $\mathcal{N}_{7}$ for
$n \geq 3$, whose $K$-matrix has the following block structure
\begin{equation}
K^{-}(x)=\pmatrix{
k_{1,1}(x) \mathbb{I}_{m \times m} & \mathbb{O}_{m\times 2n} & \mathbb{O}_{m\times m} \cr 
\mathbb{O}_{2n \times m} &  
\begin{array}{ccc}
k_{m+1,m+1}(x) &  \cdots & k_{m+1,2n+m}(x) \\ 
\vdots &  \ddots & \vdots \\ 
k_{2n+m,m+1}(x) & \cdots & k_{2n+m,2n+m}(x)
\end{array} & \mathbb{O}_{2n \times m} \cr 
\mathbb{O}_{m\times m} & \mathbb{O}_{m\times 2n} & k_{N,N}(x) \mathbb{I}_{m \times m} \cr}.
\end{equation}

The central block matrix cluster all non-diagonal elements different from zero. Concerning that central
block, we have the following expressions determining entries of the borders,
\begin{eqnarray}
k_{\alpha, 2n+m}(x) &=& \frac{\kappa }{\sqrt{\zeta }} q^{t_{\alpha} - t_{m+1} } \Omega_{m+1, \alpha^{\prime}} x G(x)
\;\;\;\;\;\;\;\;\;\;\;\;\;\;\;\;\;\;\;\;\;\;\;\;\;\;\;\; \alpha = m+2, \dots , 2n+m-1 \nonumber  \\
k_{2n+m, \alpha}(x) &=& \frac{\kappa}{\sqrt{\zeta}} q^{t_{2n+m} - t_{m+2} }
\frac{\Omega_{m+2,m+1} \Omega_{m+1,\alpha}}{\Omega_{m+1,2n+m-1}} x G(x)
\;\;\;\;\;\;\;\; \alpha =m+2, \dots ,2n+m-1 \nonumber  \\
k_{\alpha,m+1}(x) &=& q^{t_{\alpha} - t_{m+2} } \frac{\Omega_{m+2,m+1} \Omega_{m+1,\alpha^{\prime}}}{\Omega_{m+1,2n+m-1}} G(x)
\;\;\;\;\;\;\;\;\;\;\;\;\;\;\;\;\;\;\;\; \alpha=m+3, \dots ,2n+m-1 \nonumber  \\
k_{m+1 , \alpha}(x) &=& \Omega_{m+1, \alpha} G(x) 
\;\;\;\;\;\;\;\;\;\;\;\;\;\;\;\;\;\;\;\;\;\;\;\;\;\;\;\;\;\;\;\;\;\;\;\;\;\;\;\;\;\;\;\;\;\;\;\;\;\;\; \alpha=m+2, \dots ,2n+m-1
\end{eqnarray}
with $G(x)=x-1$. The entries of the secondary diagonal are given by
\begin{equation}
k_{\alpha, \alpha^{\prime}} (x) = \cases{
\displaystyle \Omega_{m+1, 2n+m} G(x) H_{b}(x) \;\;\;\;\;\;\;\;\;\;\;\;\;\;\;\;\;\;\;\;\;\;\;\;\;\;\;\;\;\;\;\;\;\;\;\;\;\;\;\;\;\;\;\;\;\;\;\;\;\;\;\;\;\;\;  \alpha = m+1 \cr
\displaystyle \frac{q^{2n-2}}{\zeta } q^{ t_{m+1} - t_{\alpha^{\prime}}}
\left( \frac{q + \kappa \sqrt{\zeta}}{q+1} \right)^{2} \frac{\Omega_{m+1,\alpha^{\prime }}^{2}}{\Omega_{m+1,2n+m}} G(x) H_{b}(x) \;\;\;\;\;\;\;\;\;
\alpha=m+2, \dots ,2n+m-1 \cr
\displaystyle q^{t_{2n+m-1} - t_{m+2} } \frac{\Omega_{m+2,m+1}^{2} \Omega_{m+1,2n+m}}{ \Omega_{m+1,2n+m-1}^{2}} G(x) H_{b}(x) 
\;\;\;\;\;\;\;\;\;\;\;\;\;\;\;\;\;\;  \alpha = 2n +m \cr}
\end{equation}
with $H_{b}(x) = \frac{xq+\kappa \sqrt{\zeta}}{q+\kappa \sqrt{\zeta}}$, and the remaining non-diagonal elements are 
determined by the expression
\begin{equation}
k_{\alpha , \beta}(x) =\cases{
\displaystyle \frac{\kappa}{\sqrt{\zeta}} q^{t_{\alpha} - t_{m+1} } \left(\frac{q+\kappa \sqrt{\zeta}}{q+1}\right)
\frac{\Omega_{m+1,\alpha^{\prime}} \Omega_{m+1,\beta}}{\beta_{m+1,2n+m}} G(x)
\;\;\;\;\;\;\;\;\;\;\;\;
\alpha < \beta^{\prime}  \; , \;  m+1 < \alpha ,\beta < 2n+m \cr
\displaystyle \frac{1}{\zeta} q^{t_{\alpha} - t_{m+1} }  \left( \frac{q+\kappa \sqrt{\zeta}}{q+1} \right) 
\frac{\Omega_{m+1,\alpha^{\prime}} \Omega_{m+1,\beta}}{\Omega_{m+1,2n+m}} x G(x)
\;\;\;\;\;\;\;\;\;\;\;\;\;
\alpha > \beta^{\prime} \; , \; m+1 < \alpha ,\beta < 2n+m \cr
\displaystyle \frac{2\kappa \sqrt{\zeta}}{(1-\kappa \sqrt{\zeta})}
\frac{(-1)^{n} \zeta (1+q)^{2}}{(q^{n}-(-1)^{n}\kappa \sqrt{\zeta})(q^{n-1}-(-1)^{n}
\kappa \sqrt{\zeta})(q+\kappa \sqrt{\zeta})}\frac{\Omega_{m+1,2n + m}}{\Omega_{m+1,m+n+1}} G(x) \cr
\;\;\;\;\;\;\;\;\;\;\;\;\;\;\;\;\;\;\;\;\;\;\;\;\;\;\;\;\;\;\;\;\;\;\;\;\;\;\;\;
\;\;\;\;\;\;\;\;\;\;\;\;\;\;\;\;\;\;\;\;\;\;\;\;\;\;\;\;\;\;\;\;\;\;\;\;\;\;\;\;\;\;\;\;\;\;\;\;\;\;\;\;\;\;\;\;\;
\alpha = m+1 \; , \; \beta=m+n \cr
\displaystyle -\frac{2\kappa \sqrt{\zeta}}{(1-\kappa \sqrt{\zeta})} \frac{q(q+\kappa \sqrt{\zeta})}{(q^{n}-(-1)^{n}\kappa \sqrt{\zeta}
)(q^{n-1}-(-1)^{n}\kappa \sqrt{\zeta})}\frac{\Omega_{m+1,m+2n-1}}{\Omega_{m+1,m+2n}} G(x) \cr
\;\;\;\;\;\;\;\;\;\;\;\;\;\;\;\;\;\;\;\;\;\;\;\;\;\;\;\;\;\;\;\;\;\;\;\;\;\;\;\;
\;\;\;\;\;\;\;\;\;\;\;\;\;\;\;\;\;\;\;\;\;\;\;\;\;\;\;\;\;\;\;\;\;\;\;\;\;\;\;\;\;\;\;\;\;\;\;\;\;\;\;\;\;\;\;\;\;
\alpha = m+2 \; , \; \beta=m+1 \cr}.
\end{equation}

In their turn the diagonal entries $k_{\alpha ,\alpha}(x)$  are given by
\begin{equation}
k_{\alpha , \alpha}(x) = \cases{
\displaystyle \left( \frac{x-\kappa \sqrt{\zeta }}{1-\kappa \sqrt{\zeta }}
\right) \left( \frac{x q^{n}-(-1)^{n}\kappa \sqrt{\zeta }}{q^{n}-(-1)^{n}\kappa \sqrt{\zeta }}\right) \left( \frac{x q^{n-1}-(-1)^{n}
\kappa \sqrt{\zeta }}{q^{n-1}-(-1)^{n}\kappa \sqrt{\zeta }}\right) \frac{2}{x(x+1)} 
\;\;\;\;\;\;\;\;\;\;\; \alpha = 1 \cr
k_{1,1}(x) + \Gamma_{n} (x)
\;\;\;\;\;\;\;\;\;\;\;\;\;\;\;\;\;\;\;\;\;\;\;\;\;\;\;\;\;\;\;\;\;\;\;\;\;\;\;\;\;\;\;\;\;\;\;\;\;\;\;\;\;\;\;\;\;\;\;\;\;\;\;\;\;\;\;\;\;\;\;\;\;\;\;\;\;\;\;\;\;\;\;\;\;\;\;\;\;\;\; \alpha = m+1 \cr
k_{m+1,m+1}(x) + \left( \Omega_{\alpha ,\alpha} - \Omega_{m+1,m+1} \right) G(x)
\;\;\;\;\;\;\;\;\;\;\;\;\;\;\;\;\;\;\;\;\;\;\;\;\;\;\; \alpha = m+2, \dots , m+n \cr
k_{n+m,n+m}(x) \;\;\;\;\;\;\;\;\;\;\;\;\;\;\;\;\;\;\;\;\;\;\;\;\;\;\;\;\;\;\;\;\;\;\;\;\;\;\;\;\;\;\;\;\;\;\;\;\;\;\;\;\;\;\;\;\;\;\;\;\;\;\;\;\;\;\;\;\; \alpha = n+m+1 \cr
k_{n+m,n+m}(x) + \left( \Omega_{\alpha ,\alpha} - \Omega_{n+m,n+m} \right) x G(x) 
\;\;\;\;\;\;\;\;\;\;\;\;\;\;\;\;\;\;\;\;\;\;\;\; \alpha = n+m+2 , \dots , 2n+m \cr
x^2 k_{1,1} (x) \;\;\;\;\;\;\;\;\;\;\;\;\;\;\;\;\;\;\;\;\;\;\;\;\;\;\;\;\;\;\;\;\;\;\;\;\;\;\;\;\;\;\;\;\;\;\;\;\;\;\;\;\;\;\;\;\;\;\;\;\;\;\;\;\;\;\;\;\;\;\;\;\;\;\;\; \alpha = N \cr}
\end{equation}
where 
\begin{equation}
\Gamma_{n}(x)=-\frac{2\zeta (q x+\kappa \sqrt{\zeta })}{(1-\kappa \sqrt{\zeta}
)(q^{n}-(-1)^{n} \kappa \sqrt{\zeta })(q^{n-1}-(-1)^{n}\kappa \sqrt{\zeta})}
\frac{G(x)}{x} .
\end{equation}

The diagonal parameters $\Omega_{\alpha, \alpha}$ are then fixed by the relations
\begin{equation}
\Omega_{\alpha , \alpha} = \cases{
\displaystyle -\frac{2\zeta (q+\kappa \sqrt{\zeta })}{
(1-\kappa \sqrt{\zeta })(q^{n}-(-1)^{n}\kappa \sqrt{\zeta }
)(q^{n-1}-(-1)^{n}\kappa \sqrt{\zeta })}
\;\;\;\;\;\;\;\;\;\;\;\;\;\;\;\;\;\; \alpha = m+1 \cr
\displaystyle \Omega_{m+1,m+1} + \Delta_{n} \sum_{k=0}^{\alpha-m-2} (-q)^{k}
\;\;\;\;\;\;\;\;\;\;\;\;\;\;\;\;\;\;\;\;\;\;\;\;\;\;\;\;\;\;\;\;\;\;\;\;\;\;\;\;\;\;\;\;\;\;\;\;\;\;\; \alpha = m+2 , \dots , n+m \cr
\displaystyle \Omega_{n+m,n+m} \;\;\;\;\;\;\;\;\;\;\;\;\;\;\;\;\;\;\;\;\;\;\;\;\;\;\;\;\;\;\;\;\;\;\;\;\;\;\;\;\;\;\;\;\;\;\;\;\;\;\;\;\;\;\;\;\;\;\;\;\;\;\;\;\;\;\;\;\;\;\;\;\;\;\;\;\;\;\; \alpha = n+m+1 \cr
\displaystyle \Omega_{n+m,n+m} + i \kappa (-1)^{n} q^{m-1} \Delta_{n} \sum_{k=0}^{\alpha-n-m-2} (-q)^{k}
\;\;\;\;\;\;\;\;\;\;\;\;\;\;\;\;\;\;\;\;\;\;\;\;\;\;
\alpha = n+m+2 , \dots , 2n+m  \cr}
\end{equation}
and the auxiliary parameter $\Delta_{n}$ is given by 
\begin{equation}
\Delta_{n}=\frac{2\zeta (1+q)^{2}}{(1-\kappa \sqrt{\zeta })(q^{n}-(-1)^{n}
\kappa \sqrt{\zeta })(q^{n-1}-(-1)^{n}\kappa \sqrt{\zeta })} .
\end{equation}
Besides the above relations the following constraints should also holds 
\begin{eqnarray}
\Omega_{m+1,m+\alpha} &=& - \Omega_{m+1,\alpha +m+1} \frac{\Omega_{m+1,2n + m - \alpha}}{\Omega_{m+1,2n + m + 1  - \alpha}} \;\;\;\;\;\;\;\;\;\;\;\;\;\;\;\;\;\;\;\;\;\; \alpha = 2, \dots , n-1 ,
\end{eqnarray}
and $\Omega_{m+1, m+n+1} , \dots , \Omega_{m+1 , 2n+m}$ are regarded as free parameters.

\vskip 1cm
\begin{itemize}
\item {\bf Solution $\mathcal{N}_{8}$:}
\end{itemize}
For $n \geq 2$ the vertex model $U_{q}[sl^{(2)}(2n+1|2m)]$ admits the family of solutions $\mathcal{N}_{8}$ whose $K$-matrix
is of the form
\begin{equation}
K^{-}(x)=\pmatrix{
k_{1,1}(x) \mathbb{I}_{m \times m} & \mathbb{O}_{m\times (2n+1)} & \mathbb{O}_{m\times m} \cr 
\mathbb{O}_{(2n+1) \times m} &  
\begin{array}{ccc}
k_{m+1,m+1}(x) &  \cdots & k_{m+1,2n+m+1}(x) \\ 
\vdots &  \ddots & \vdots \\ 
k_{2n+m+1,m+1}(x) & \cdots & k_{2n+m+1+,2n+m+1}(x)
\end{array} & \mathbb{O}_{(2n+1) \times m} \cr 
\mathbb{O}_{m\times m} & \mathbb{O}_{m\times (2n+1)} & k_{N,N}(x) \mathbb{I}_{m \times m} \cr}.
\end{equation}
In the central block matrix we find all non-diagonal elements different from zero similarly to the structure of the solution
$\mathcal{N}_{7}$. The borders of the central block are then determined by the following expressions
\begin{eqnarray}
k_{\alpha, 2n+m+1}(x) &=& \frac{\kappa }{\sqrt{\zeta }} q^{t_{\alpha} - t_{m+1} } \Omega_{m+1, \alpha^{\prime}} x G(x)
\;\;\;\;\;\;\;\;\;\;\;\;\;\;\;\;\;\;\;\;\;\;\;\;\;\;\;\;\;\;\; \alpha = m+2, \dots , 2n+m \nonumber \\
k_{2n+m+1, \alpha}(x) &=& \frac{\kappa}{\sqrt{\zeta}} q^{t_{2n+m+1} - t_{m+2} }
\frac{\Omega_{m+2,m+1} \Omega_{m+1,\alpha}}{\Omega_{m+1,2n+m}} x G(x)
\;\;\;\;\;\;\;\; \alpha =m+2, \dots ,2n+m \nonumber \\
k_{\alpha,m+1}(x) &=& q^{t_{\alpha} - t_{m+2} } \frac{\Omega_{m+2,m+1} \Omega_{m+1,\alpha^{\prime}}}{\Omega_{m+1,2n+m}} G(x)
\;\;\;\;\;\;\;\;\;\;\;\;\;\;\;\;\;\;\;\;\;\;\; \alpha=m+3, \dots ,2n+m \nonumber \\
k_{m+1 , \alpha}(x) &=& \Omega_{m+1, \alpha} G(x) 
\;\;\;\;\;\;\;\;\;\;\;\;\;\;\;\;\;\;\;\;\;\;\;\;\;\;\;\;\;\;\;\;\;\;\;\;\;\;\;\;\;\;\;\;\;\;\;\;\;\;\;\;\;\; \alpha=m+2, \dots ,2n+m
\end{eqnarray}
recalling that $\kappa = \pm 1$ and $G(x)=x-1$. The secondary diagonal elements are given by
\begin{equation}
k_{\alpha, \alpha^{\prime}} (x) = \cases{
\displaystyle \Omega_{m+1, 2n+m+1} G(x) H_{b}(x) \;\;\;\;\;\;\;\;\;\;\;\;\;\;\;\;\;\;\;\;\;\;\;\;\;\;\;\;\;\;\;\;\;\;\;\;\;\;\;\;\;\;\;\;\;\;\;\;\;\;\;\;\;\;\;  \alpha = m+1 \cr
\displaystyle \frac{q^{2n-1}}{\zeta } q^{ t_{m+1} - t_{\alpha^{\prime}}}
\left( \frac{q + \kappa \sqrt{\zeta}}{q+1} \right)^{2} \frac{\Omega_{m+1,\alpha^{\prime }}^{2}}{\Omega_{m+1,2n+m+1}} G(x) H_{b}(x) \;\;\;\;\;\;\;\;\;
\alpha=m+2, \dots ,2n+m \cr
\displaystyle q^{t_{2n+m} - t_{m+2} } \frac{\Omega_{m+2,m+1}^{2} \Omega_{m+1,2n+m+1}}{ \Omega_{m+1,2n+m}^{2}} G(x) H_{b}(x) 
\;\;\;\;\;\;\;\;\;\;\;\;\;\;\;\;\;\;\;\;\;  \alpha = 2n + m +1 \cr}
\end{equation}
where $H_{b}(x) = \frac{xq+\kappa \sqrt{\zeta}}{q+\kappa \sqrt{\zeta}}$.

The remaining non-diagonal entries are determined by 
\begin{equation}
k_{\alpha , \beta}(x) = \cases{
\displaystyle \frac{\kappa}{\sqrt{\zeta}} q^{t_{\alpha} - t_{m+1} } \left( 
\frac{q+\kappa \sqrt{\zeta}}{q+1}\right) \frac{\Omega_{m+1, \alpha^{\prime}} \Omega_{m+1,\beta}}{\Omega_{m+1,2n+m+1}} G(x)
\;\;\;\;\;\;\;\;\; \alpha < \beta^{\prime} \; , \; m+2 \leq \alpha , \beta \leq 2n+m \cr
\displaystyle \frac{1}{\zeta} q^{t_{\alpha} - t_{m+1} } \left( \frac{q+\kappa \sqrt{\zeta}}{q+1}\right) \frac{\Omega_{m+1,\alpha^{\prime }}\Omega_{m+1,\beta}}{\Omega_{m+1,2n+m+1}} x G(x)
\;\;\;\;\;\;\;\;\;\; \alpha > \beta^{\prime} \; , \; m+2 \leq \alpha , \beta \leq 2n+m \cr
\displaystyle (-1)^n q^{2m-2n} \left(\frac{q + \kappa \sqrt{\zeta} }{q+1} \right)^2 \frac{\Omega_{m+1,n+m} \Omega_{m+1,n+m+2} \Omega_{m+1,2n+m}}{\Omega_{m+1,2n+m+1}^2} G(x) \cr
\;\;\;\;\;\;\;\;\;\;\;\;\;\;\;\;\;\;\;\;\;\;\;\;\;\;\;\;\;\;\;\;\;\;\;\;\;\;\;\;\;\;\;\;\;\;\;\;\;\;\;\;\;\;\;\;\;\;\;\;\;\;\;\;\;\;\;\;\;\;\;\;\;\;\;\;\;\;\;\;\;\;\;\;\;\;\;\;\;\;\;\;\;\;\; \alpha = m+2 \; , \; \beta = m+1 \cr
\displaystyle \left[ \frac{i \kappa (-1)^n q^{m-1} (q+1) }{(q^{m-\frac{1}{2}} - i\kappa (-1)^n )^2}\frac{\Omega_{m+1,n+m+1}^2}{\Omega_{m+1,n+m+2}} \right. \cr
\displaystyle \left. \;\; - \frac{2 \kappa (-1)^n(q+1)^2 \sqrt{\zeta}} {(q^{m-\frac{1}{2}} - i \kappa (-1)^n)^2 (1 - \kappa \sqrt{\zeta}) (q + \kappa \sqrt{\zeta})} \frac{\Omega_{m+1,2n+m+1}}{\Omega_{m+1,m+n+2}} \right] G(x) \cr
\;\;\;\;\;\;\;\;\;\;\;\;\;\;\;\;\;\;\;\;\;\;\;\;\;\;\;\;\;\;\;\;\;\;\;\;\;\;\;\;\;\;\;\;\;\;\;\;\;\;\;\;\;\;\;\;\;\;\;\;\;\;\;\;\;\;\;\;\;\;\;\;\;\;\;\;\;\;\;\;\;\;\;\;\;\;\;\;\;\;\;\;\;\;\; \alpha = m+1 \; , \; \beta = m+n \cr},
\end{equation}
while the diagonal matrix elements are given by the relations
\begin{equation}
k_{\alpha , \alpha}(x) = \cases{
\displaystyle \left[ \frac{2 (x - \kappa \sqrt{\zeta}) (x q^{m-\frac{1}{2}} - i \kappa (-1)^{n})^2 }{x (x^2-1) (1- \kappa \sqrt{\zeta}) (q^{m-\frac{1}{2}} - i \kappa (-1)^n)^2} \right. \cr
\displaystyle \;\;\; - \left. \frac{i q^{m-1} (1+q^{2m-1} x) ( q + \kappa \sqrt{\zeta}) (x - \kappa \sqrt{\zeta})}  {\sqrt{\zeta} x (x+1) (q+1) (q^{m-\frac{1}{2}} - i \kappa (-1)^n)^2} \right] G(x)
\;\;\;\;\;\;\;\;\;\;\;\;\;\;\;\;\;\;\; \alpha = 1 \cr
k_{1,1}(x) + \Gamma(x) \;\;\;\;\;\;\;\;\;\;\;\;\;\;\;\;\;\;\;\;\;\;\;\;\;\;\;\;\;\;\;\;\;\;\;\;\;\;\;\;\;\;\;\;\;\;\;\;\;\;\;\;\;\;\;\;\;\;\;\;\;\;\;\;\;\;\;\;\;\;\;\;\;\;\;\;\; \alpha = m+1 \cr
k_{m+1,m+1}(x) + \left( \Omega_{\alpha , \alpha} - \Omega_{m+1,m+1} \right) G(x)
\;\;\;\;\;\;\;\;\;\;\;\;\;\;\;\;\;\;\;\;\;\;\;\;\;\;\;\;\;\;\;\;\;\;\;\;\; \alpha = m+2, \dots , m+n \cr
k_{m+1,m+1}(x) + \left( \Omega_{n+m+1,n+m+1} - \Omega_{m+1,m+1} \right)G(x) + \Delta (x)
\;\;\;\;\;\;\;\;\;\;\;\;\;\;\;\;\;\;\;\; \alpha = n+m+1 \cr
k_{n+m+1,n+m+1}(x) + \left( \Omega_{n+m+2,n+m+2} - \Omega_{n+m+1,n+m+1} \right) x G(x) + \kappa \sqrt{\zeta} \Delta(x) \cr
\;\;\;\;\;\;\;\;\;\;\;\;\;\;\;\;\;\;\;\;\;\;\;\;\;\;\;\;\;\;\;\;\;\;\;\;\;\;\;\;\;\;\;\;\;\;\;\;\;\;\;\;\;\;\;\;\;\;\;\;\;\;\;\;
\;\;\;\;\;\;\;\;\;\;\;\;\;\;\;\;\;\;\;\;\;\;\;\;\;\;\;\;\;\;\;\;\;\;\;\;\;\;\;\;\;\;\;\;\;\;\; \alpha = n+m+2 \cr
k_{\alpha-1 ,\alpha-1}(x) + \left( \Omega_{\alpha ,\alpha} - \Omega_{\alpha-1,\alpha-1} \right) x G(x)
\;\;\;\;\;\;\;\;\;\;\;\;\;\;\;\;\;\;\;\;\;\;\;\; \alpha = n+m+3 , \dots , 2n + m +1 \cr
x^2 k_{1,1}(x) \;\;\;\;\;\;\;\;\;\;\;\;\;\;\;\;\;\;\;\;\;\;\;\;\;\;\;\;\;\;\;\;\;\;\;\;\;\;\;\;\;\;\;\;\;\;\;\;\;\;\;\;\;\;\;\;\;\;\;\;\;\;\;\;\;\;\;\;\;\;\;\; \alpha = N \cr}.
\end{equation}
The auxiliary functions $\Delta(x)$ and $\Gamma(x)$  are
\begin{eqnarray}
\Delta(x) &=& q^{2m-n-1} \frac{(q+\kappa \sqrt{\zeta})}{(q+1)^2}   \frac{\Omega_{m+1,n+m} \Omega_{m+1,n+m+2} }{\Omega_{m+1,2n+m+1}}
(x-1)^2 \\
\Gamma(x) &=& (-1)^{n+1} \frac{(x q + \kappa \sqrt{\zeta}) (q + \kappa \sqrt{\zeta})}{(\kappa \sqrt{\zeta} x (q+1)^2)}
\frac{ \Omega_{m+1,n+m} \Omega_{m+1,n+m+2}}{\Omega_{m+1,2n+m+1}} G(x) ,
\end{eqnarray}
and the parameters $\Omega_{m+1,m+\alpha}$ are constrained by the recurrence formula
\begin{equation}
\Omega_{m+1,\alpha + m} = - \frac{\Omega_{m+1,\alpha+ m+1} \Omega_{m+1,2n +m+1-\alpha} }{\Omega_{m+1,2n+m+2-\alpha}}
\;\;\;\;\;\;\;\;\;\;\;\;\;\;\; \alpha = 2, \dots , n-1 .
\end{equation}
In their turn the diagonal parameters $\Omega_{\alpha ,\alpha}$ are fixed by
\begin{equation}
\Omega_{\alpha ,\alpha}=\cases{
\displaystyle \frac{2 i \kappa (q+\kappa \sqrt{\zeta})}{(q^{n-m+\frac{1}{2}} + i \kappa) (q^{m - \frac{1}{2}} - i \kappa (-1)^n)^2}  
- \frac{q^{2m-n-\frac{3}{2}} (q + \kappa \sqrt{\zeta})^2 }{(q+1) (q^{m-\frac{1}{2}} - i \kappa (-1)^n)^2} \frac{\Omega_{m+1,n+m+1}^2}{\Omega_{m+1,2n+m+1}} \cr
\;\;\;\;\;\;\;\;\;\;\;\;\;\;\;\;\;\;\;\;\;\;\;\;\;\;\;\;\;\;\;\;\;\;\;\;\;\;\;\;\;\;\;\;\;\;\;\;\;\;\;\;\;\;\;\;\;\;\;\;\;\;\;\;\;\;\;\;\;\;\;\;\;\;\;\;\;\;\;\;\;\;\;\;\;\;\;\;\;\;\;\;\;\;\;\;\;\;\;\;\;\;\;\;\;\;\;\;\;\;\;\;\;\;\;\; \alpha = m+1 \cr
\displaystyle \Omega_{m+1,m+1} + Q_{n,m} \sum_{k=0}^{\alpha-m-2} (-q)^k 
\;\;\;\;\;\;\;\;\;\;\;\;\;\;\;\;\;\;\;\;\;\;\;\;\;\;\;\;\;\;\;\;\;\;\;\;\;\;\;\;\;\;\;\;\;\;\;\;\;\;\; \alpha = m+2, \dots , n+m \cr
\displaystyle \frac{(q^{m-1} - i \kappa q^{2m-n-\frac{1}{2}}) (q^n-(-1)^n) }{(q+1)(q^{m-\frac{1}{2}} - i \kappa (-1)^n)}\frac{\Omega_{m+1,n+m+1}^2}{\Omega_{m+1,2n+m+1}} - \frac{2 i \kappa (-1)^n }{(q^{m-\frac{1}{2}} -i \kappa (-1)^n)} \cr
\;\;\;\;\;\;\;\;\;\;\;\;\;\;\;\;\;\;\;\;\;\;\;\;\;\;\;\;\;\;\;\;\;\;\;\;\;\;\;\;\;\;\;\;\;\;\;\;\;\;\;\;\;\;\;\;\;\;\;\;\;\;\;\;\;\;\;\;\;\;\;\;\;\;\;\;\;\;\;\;\;\;\;\;\;\;\;\;\;\;\;\;\;\;\;\;\;\;\;\;\;\;\;\;\;\;\;\;\;\; \alpha = n+m+1 \cr
\displaystyle \Omega_{n+m+1,n+m+1} - q^{2m-n-1} \left( \frac{q + \kappa \sqrt{\zeta} }{q+1} \right)^2 \frac{\Omega_{m+1,n+m} \Omega_{m+1,n+m+2}}{\Omega_{m+1,2n+m+1}} \cr
\displaystyle +  q^{2m-n-\frac{3}{2}} \frac{q+\kappa \sqrt{\zeta} }{q+1} \frac{\Omega_{m+1,n+m+1}^2}{\Omega_{m+1,2n+m+1}}  \;\;\;\;\;\;\;\;\;\;\;\;\;\;\;\;\;\;\;\;\;\;\;\;\;\;\;\;\;\;\;\;\;\;\;\;\;\;\;\;\;\;\;\;\;\;\;\;\;\;\;\;\;\;\;\;\;\; \alpha = n+m+2 \cr
\displaystyle  \Omega_{n+m+2,n+m+2} + (-1)^n \kappa  q^{2m-n-1} \sqrt{\zeta}  Q_{n,m} \sum_{k=0}^{\alpha - n-m-3} (-q)^k \cr
\;\;\;\;\;\;\;\;\;\;\;\;\;\;\;\;\;\;\;\;\;\;\;\;\;\;\;\;\;\;\;\;\;\;\;\;\;\;\;\;\;\;\;\;\;\;\;\;\;\;\;\;\;\;\;\;\;\;\;\;\;\;\;\;\;\;\;\;\;\;\;\;\;\;\;\;\;\;\;\;\;\;\;\;\; \alpha = n+m+3 , \dots, 2n+m+1 \cr}
\end{equation}
where
\begin{equation}
Q_{n,m} = \frac{2 (q+1)^2}{(\kappa \sqrt{\zeta} - 1 )(q^{m-\frac{1}{2}} - i \kappa (-1)^n)^2 }
+ i q^{m-1} \frac{(q+1)(q + \kappa \sqrt{\zeta}) }{\sqrt{\zeta} (q^{m-\frac{1}{2}}- i \kappa (-1)^n)^2}
\frac{\Omega_{m+1,n+m+1}^2}{\Omega_{m+1,2n+m+1}} .
\end{equation}

This solution has altogether $n+1$ free parameters corresponding to the set of variables $\Omega_{m+1, n+m+1} , \dots , \Omega_{m+1,2n+m+1}$.

\vskip 1cm
\begin{itemize}
\item {\bf Solution $\mathcal{N}_{9}$:}
\end{itemize}

The family $\mathcal{N}_{9}$ consist of a solution of the reflection equation where all entries of the $K$-matrix are
non-null. This solution is admitted only by the $U_{q}[sl(1|2m)^{(2)}]$ vertex model. The associated $K$-matrix is of the general form
(\ref{KM}) and the matrix elements of the borders are mainly given by
\begin{eqnarray}
k_{\alpha,N}(x)&=&-\frac{\kappa}{\sqrt{\zeta}}\frac{\epsilon_{\alpha}}{\epsilon_{1}}
q^{t_{\alpha}-t_{1}}\Omega_{1,\alpha^{\prime}} x G(x) \;\;\;\;\;\;\;\;\;\;\;\;\;\;\;\;\;\;\;\;\; \alpha = 2, \dots , N-1 \nonumber \\
k_{\alpha,1}(x) &=& \frac{\epsilon_{\alpha}}{\epsilon_{2}} q^{t_{\alpha}-t_{2}}\frac{\Omega_{2,1} \Omega_{1, \alpha^{\prime }}}{\Omega_{1,N-1}} G(x) \;\;\;\;\;\;\;\;\;\;\;\;\;\;\;\;\;\;\;\;\;\;\;\;\; \alpha = 3, \dots , N-1 \nonumber \\
k_{N,\alpha}(x) &=& -\frac{\kappa}{\sqrt{\zeta}}\frac{\epsilon_{N}}{\epsilon_{2}}
q^{t_{N}-t_{2}}\frac{\Omega_{2,1} \Omega_{1,\alpha}}{\Omega_{1,N-1}} x G(x)
\;\;\;\;\;\;\;\;\;\;\;\;\;\; \alpha = 2, \dots , N-1 \nonumber \\
k_{1,\alpha}(x) &=& \Omega_{1, \alpha} G(x) 
\;\;\;\;\;\;\;\;\;\;\;\;\;\;\;\;\;\;\;\;\;\;\;\;\;\;\;\;\;\;\;\;\;\;\;\;\;\;\;\;\;\;\;\;\;\; \alpha = 2, \dots , N-1.
\end{eqnarray}

The secondary diagonal is characterized by entries of the form
\begin{equation}
k_{\alpha ,\alpha'}(x)=\cases{
\displaystyle \Omega_{1,N} G(x) H_{f}(x) \;\;\;\;\;\;\;\;\;\;\;\;\;\;\;\;\;\;\;\;\;\;\;\;\;\;\;\;\;\;\;\;\;\;\;\;\;\;\;\;\;\;\;\;\;\;\;\;\;\;\;\;\;\;\; \alpha = 1 \cr
\displaystyle q^{t_{1}-t_{\alpha^{\prime}}} \frac{\epsilon_{1}}{\epsilon_{\alpha^{\prime}}} \left( \frac{1-\kappa q\sqrt{\zeta} }{q+1} \right)^2 \frac{\Omega_{1,\alpha^{\prime}}^{2}}{\Omega_{1,N}} G(x) H_{f}(x)  \;\;\;\;\;\;\;\;\;\;\;\;\;\;\; \alpha = 2, \dots , m  \;\; \mbox{and} \cr
\;\;\;\;\;\;\;\;\;\;\;\;\;\;\;\;\;\;\;\;\;\;\;\;\;\;\;\;\;\;\;\;\;\;\;\;\;\;\;\;\;\;\;\;\;\;\;\;\;\;\;\;\;\;\;\;\;\;\;\;\;\;\;\;\;\;\;\;\;\;\;\;\;\;\;\;\;\;\;\; \alpha = m+2, \dots , N-1 \cr
\displaystyle \frac{\epsilon_{N-1}}{\epsilon_{2}} q^{t_{N-1}-t_{2}}\frac{\Omega_{1,N} \Omega_{2,1}^{2}}{\Omega_{1,N-1}^{2}} G(x) H_{f}(x)
\;\;\;\;\;\;\;\;\;\;\;\;\;\;\;\;\;\;\;\;\;\;\;\;\;\;\; \alpha = N \cr}
\end{equation}
where $G(x)=x-1$ and $H_{f}(x)= \displaystyle \frac{x-\kappa q \sqrt{\zeta} }{1-\kappa q\sqrt{\zeta}}$, and the remaining non-diagonal elements are given by
\begin{equation}
k_{\alpha, \beta}(x) = \cases{
\displaystyle - \frac{\kappa}{\sqrt{\zeta}} \frac{\epsilon_{\alpha}}{\epsilon_{1}}
q^{t_{\alpha}-t_{1}} \left( \frac{1-\kappa q\sqrt{\zeta}}{q+1} \right) \frac{\Omega_{1,\alpha^{\prime}} \Omega_{1,\beta}}{\Omega_{1,N}} G(x) \;\;\;\;\;\;\;\;\;\;\;\;\; \alpha < \beta^{\prime} \; , \; 2 \leq \alpha ,\beta \leq N-1 \cr
\displaystyle \frac{1}{\zeta}\frac{\epsilon_{\alpha}}{\epsilon_{1}}
q^{t_{\alpha}-t_{1}} \left(\frac{1-\kappa q\sqrt{\zeta}}{q+1}\right) \frac{\Omega_{1,\alpha^{\prime}} \Omega_{1,\beta}}{\Omega_{1,N}} x G(x) \;\;\;\;\;\;\;\;\;\;\;\;\;\;\;\;\; \alpha > \beta^{\prime} \; ,\; 2 \leq \alpha ,\beta \leq N-1 \cr
\displaystyle \frac{i \sqrt{q}}{(q-1)} \frac{\Omega_{1,m+1}^2}{\Omega_{1,m+2}} G(x) \;\;\;\;\;\;\;\;\;\;\;\;\;\;\;\;\;\;\;\;\;\;\;\;\;\;\;\;\;\;\;\;\;\;\;\;\;\;\;\;\;\;\;\;\;\;\;\;\; \alpha=1 \; , \; \beta =m \cr
\displaystyle - \frac{i}{2} \Omega_{1,m+1}^2 \frac{\left(q^{\frac{3}{2}} + i \kappa (-1)^m \right) (q^{m-\frac{1}{2}} + i \kappa) ( \kappa q \sqrt{\zeta} - 1 ) }{(q^2-1) \left(\sqrt{q} - i \kappa (-1)^m \right)} G(x) \;\;\;\;\;\;\;\;\;\; \alpha=1 \; , \; \beta =N \cr
\displaystyle (-1)^m q^{2m-2} \left( \frac{q \kappa \sqrt{\zeta}-1 }{q+1} \right)^2 \frac{\Omega_{1,m} \Omega_{1,2m} \Omega_{1,m+2} }{\Omega_{1,2m+1}^2} G(x) \;\;\;\;\;\;\;\;\;\;\;\;\;\;\;\;\;\;\;\; \alpha=2 \; , \; \beta =1 \cr}.
\end{equation}

In their turn the diagonal entries $k_{\alpha , \alpha}(x)$ are given by the following expression
\begin{equation}
k_{\alpha , \alpha}(x) = \cases{
\displaystyle \left( \frac{2 x}{x^2-1} - \frac{\Omega_{m+1,m+1} }{x+1} \right) G(x)
+ \left( \frac{ 1+\kappa \sqrt{\zeta} }{x^2-1}  \right) \Gamma(x) 
\;\;\;\;\;\;\;\;\;\;\;\;\;\;\;\;\;\;\;\;\;\;\;\;\;\;\;\; \alpha =1 \cr
k_{1,1}(x) +  \Omega_{\alpha, \alpha}  G(x) 
\;\;\;\;\;\;\;\;\;\;\;\;\;\;\;\;\;\;\;\;\;\;\;\;\;\;\;\;\;\;\;\;\;\;\;\;\;\;\;\;\;\;\;\;\;\;\;\;\;\;\;\;\;\;\;\;\;\;\;\;\;\;\;\;\;\;\;\;\;\;\;\;\;\;\; \alpha = 2, \dots , m \cr
\displaystyle k_{1,1}(x) + \Omega_{m+1,m+1} G(x) + \Gamma(x)
\;\;\;\;\;\;\;\;\;\;\;\;\;\;\;\;\;\;\;\;\;\;\;\;\;\;\;\;\;\;\;\;\;\;\;\;\;\;\;\;\;\;\;\;\;\;\;\;\;\;\;\;\;\;\; \alpha =m+1 \cr
\displaystyle  k_{m+1,m+1}(x) + \left( \Omega_{m+2,m+2} - \Omega_{m+1,m+1} \right) x G(x) + i \kappa q^{\frac{1}{2}-m} \Gamma(x)
\;\;\;\;\;\;\;\;\; \alpha =m+2 \cr
\displaystyle  k_{\alpha-1,\alpha-1}(x) + \left( \Omega_{\alpha, \alpha} - \Omega_{\alpha - 1,\alpha - 1} \right) x G(x)
\;\;\;\;\;\;\;\;\;\;\;\;\;\;\;\;\;\;\;\;\;\;\;\;\;\;\;\;\;\;\;\;\;\;\;\;\;\;\;\;\;\;\;\; \alpha = m+3 , \dots , N \cr}
\end{equation}
where the auxiliary function $\Gamma(x)$ is given by
\begin{equation}
\Gamma(x) = \frac{q^{m} (x-1)^2 ( \kappa q \sqrt{\zeta} - 1 ) }{(q+1)^2 } \frac{\Omega_{1,m} \Omega_{1,m+2} }{\Omega_{1, 2m+1}}  .
\end{equation}
The parameters $\Omega_{1 , \alpha}$ are constrained by the recurrence relation
\begin{equation}
\Omega_{1,\alpha} = - \frac{\Omega_{1,\alpha+1} \Omega_{1,2m+1-\alpha}}{\Omega_{1, 2m + 2 - \alpha}}
\;\;\;\;\;\;\;\;\;\;\;\;\;\;\;\;\;\;\;\;\; \alpha = 2 , \dots , m-1 \; 
\end{equation}
while $\Omega_{\alpha , \alpha}$ are fixed by 
\begin{equation}
\Omega_{\alpha ,\alpha} = \cases{
\displaystyle Q_{m} \sum_{k=0}^{\alpha -2} (-\frac{1}{q})^k
\;\;\;\;\;\;\;\;\;\;\;\;\;\;\;\;\;\;\;\;\;\;\;\;\;\;\;\;\;\;\;\;\;\;\;\;\;\;\;\;\;\;\;\;\;\;\;\;\;\;\;\;\;\;\;\;\;\;\;\;\;\;\;\;\;\;\; \alpha = 2, \dots , m \cr
\displaystyle \Omega_{m,m} + \delta_{1} 
\;\;\;\;\;\;\;\;\;\;\;\;\;\;\;\;\;\;\;\;\;\;\;\;\;\;\;\;\;\;\;\;\;\;\;\; \;\;\;\;\;\;\;\;\;\;\;\;\;\;\;\;\;\; \;\;\;\;\;\;\;\;\;\;\;\;\;\;\;\;\;\; \; \alpha = m+1 \cr
\displaystyle \Omega_{m,m} + \delta_{2}
\;\;\;\;\;\;\;\;\;\;\;\;\;\;\;\;\;\;\;\;\;\;\;\;\;\;\;\;\;\;\;\;\;\;\;\; \;\;\;\;\;\;\;\;\;\;\;\;\;\;\;\;\;\; \;\;\;\;\;\;\;\;\;\;\;\;\;\;\;\;\;\; \; \alpha = m+2 \cr
\displaystyle \Omega_{m+2,m+2} + (-1)^m q^{m-1} \kappa \sqrt{\zeta} Q_{m} \sum_{k=0}^{\alpha -m-3}  (-\frac{1}{q})^k 
\;\;\;\;\;\;\;\;\;\;\;\;\;\;\;\;\;\; \alpha = m+3 , \dots , N-1 \cr
\displaystyle 2  \;\;\;\;\;\;\;\;\;\;\;\;\;\;\;\;\;\;\;\;\;\;\;\;\;\;\;\;\;\;\;\;\;\;\;\; \;\;\;\;\;\;\;\;\;\;\;\;\;\;\;\;\;\; \;\;\;\;\;\;\;\;\;\;\;\;\;\;\;\;\;\;\;\;\;\;\;\;\;\;\;\;\; \;\;\;\;\; \alpha = N \cr}
\end{equation}
where
\begin{eqnarray}
Q_{m} &=& 2 i \kappa (-1)^m  q^{m-1}  \frac{(q+1) \left( \sqrt{q} - i \kappa (-1)^m \right)}{( q^{m-\frac{1}{2}}+ i \kappa ) ( q^{\frac{3}{2}} + i \kappa (-1)^m )} , \nonumber \\
\delta_{1} &=& -2 \frac{(q^{m+\frac{1}{2}} + 2 i \kappa q + i \kappa )}{(\sqrt{q} + i \kappa (-1)^m ) (q^{\frac{3}{2}} + i \kappa (-1)^m ) ( q^{m-\frac{1}{2}} + i \kappa )} ,  \\
\delta_{2} &=& 2 \frac{(q^{m+\frac{1}{2}}- i \kappa) \left(\sqrt{q} - i \kappa (-1)^m \right)}{( q^{\frac{3}{2}} + i \kappa (-1)^m ) ( q^{m-\frac{1}{2}} + i \kappa )}. \nonumber
\end{eqnarray}
In this solution the have a total amount of $m$ free parameters, namely $\Omega_{1, m+1} , \dots , \Omega_{1, 2m}$.

\vskip 1cm
\begin{itemize}
\item {\bf Solution $\mathcal{N}_{10}$:}
\end{itemize}

The series of solutions $\mathcal{N}_{10}$ is valid for the $U_{q}[sl(2|2m)^{(2)}]$ model and the corresponding $K$-matrix also possess all entries different from zero. In the first and last columns, the matrix elements are mainly given by
\begin{eqnarray}
k_{\alpha ,1}(x) &=& \frac{\epsilon_{\alpha}}{\epsilon_{2}} q^{t_{\alpha} - t_{2}} \frac{\Omega_{2,1} \Omega_{1,\alpha^{\prime}}}{\Omega_{1,N-1}} G(x) \;\;\;\;\;\;\;\;\;\;\;\;\;\;\;\;\;\;\; \alpha = 3, \dots, N-1 \nonumber \\
k_{\alpha ,N}(x) &=& -\frac{\kappa}{\sqrt{\zeta}}\frac{\epsilon_{\alpha}}{\epsilon_{1}}
q^{t_{\alpha}-t_{1}} \Omega_{1, \alpha^{\prime}} x G(x) 
\;\;\;\;\;\;\;\;\;\;\;\;\;\;\;\; \alpha = 2 , \dots , N-1
\end{eqnarray}
while the ones in the first and last rows are respectively
\begin{eqnarray}
k_{1, \alpha}(x) &=& \Omega_{1, \alpha} G(x) \;\;\;\;\;\;\;\;\;\;\;\;\;\;\;\;\;\;\;\;\;\;\;\;\;\;\;\;\;\;\;\;\;\;\;\;\;\;\;\;\;\;\;\; \alpha = 2, \dots , N-1 \nonumber \\
k_{N, \alpha}(x) &=& -\frac{\kappa}{\sqrt{\zeta}}\frac{\epsilon_{N}}{\epsilon_{2}}
q^{t_{N}-t_{2}} \frac{\Omega_{2,1} \Omega_{1,\alpha}}{\Omega_{1,N-1}} x G(x)
\;\;\;\;\;\;\;\;\;\;\;\; \alpha = 2, \dots , N-1 
\end{eqnarray}
with $G(x)=x-1$.

In the secondary diagonal we have the following expression determining the matrix elements
\begin{equation}
k_{\alpha , \alpha'}(x) = \cases{
\displaystyle \Omega_{1,N} G(x) H_{f}(x) \;\;\;\;\;\;\;\;\;\;\;\;\;\;\;\;\;\;\;\;\;\;\;\;\;\;\;\;\;\;\;\;\;\;\;\;\;\;\;\;\;\;\;\;\;\;\;\;\;\;\;\; \alpha =1 \cr
\displaystyle q^{t_{1}-t_{\alpha^{\prime}}} \frac{\epsilon_{1}}{\epsilon_{\alpha^{\prime}}} \left( \frac{1-\kappa q\sqrt{\zeta}}{q+1} \right)^2 \frac{\Omega_{1,\alpha^{\prime}}^{2}}{\Omega_{1,N}} G(x) H_{f}(x)
\;\;\;\;\;\;\;\;\;\;\;\;  \alpha = 2, \dots , m  \;\; \mbox{and} \cr
\;\;\;\;\;\;\;\;\;\;\;\;\;\;\;\;\;\;\;\;\;\;\;\;\;\;\;\;\;\;\;\;\;\;\;\;\;\;\;\;\;\;\;\;\;\;\;\;\;\;\;\;\;\;\;\;\;\;\;\;\;\;\;\;\;\;\;\;\;\;\;\;\;\;\;\;\; \alpha = m+3, \dots , N-1 \cr
\displaystyle q^{t_{1}-t_{\alpha^{\prime}}}\frac{\epsilon_{1}}{\epsilon_{\alpha^{\prime}}}\frac{(1-\kappa q\sqrt{\zeta})(q+\kappa \sqrt{\zeta})}{(q^{2}-1)}\frac{\Omega_{1,\alpha^{\prime}}^{2}}{\Omega_{1,N}} G(x) H_{b}(x)
\;\;\;\;\;\;\;\;\;\;\;\;\;\;\;\;  \alpha = m+1,m+2 \cr
\displaystyle \frac{\epsilon_{N-1}}{\epsilon_{2}} q^{t_{N-1}-t_{2}} \frac{\Omega_{1,N} \Omega_{2,1}^{2}}{\Omega_{1,N-1}^{2}} G(x) H_{f}(x)
\;\;\;\;\;\;\;\;\;\;\;\;\;\;\;\;\;\;\;\;\;\;\;\;\;\;\;\;\;\;\;\;\;\;\;\;\;\; \alpha =N \cr}
\end{equation}
recalling that 
\begin{equation}
\label{hfhb}
H_{b}(x)=\frac{q x + \kappa \sqrt{\zeta}}{q+\kappa \sqrt{\zeta}} \;\;\; \mbox{and} \;\;\;
H_{f}(x)=\frac{x-\kappa q \sqrt{\zeta}}{1-\kappa q\sqrt{\zeta}}. 
\end{equation}

In their turn the other non-diagonal entries satisfy the relation
\begin{equation}
k_{\alpha , \beta}(x) = \cases{
\displaystyle - \frac{\kappa}{\sqrt{\zeta}}\frac{\epsilon_{\alpha}}{\epsilon_{1}}
q^{t_{\alpha}-t_{1}}\left( \frac{1-\kappa q\sqrt{\zeta}}{q+1}\right) \frac{\Omega_{1,\alpha^{\prime}} \Omega_{1,\beta}}{\Omega_{1,N}} G(x)
\;\;\;\;\;\;\;\;\;\;\; \alpha < \beta^{\prime} \; , \; 2 \leq \alpha , \beta \leq N-1 \cr
\displaystyle \frac{1}{\zeta}\frac{\epsilon_{\alpha}}{\epsilon_{1}}
q^{t_{\alpha}-t_{1}}\left( \frac{1-\kappa q\sqrt{\zeta}}{q+1}\right) \frac{\Omega_{1,\alpha^{\prime}} \Omega_{1,\beta}}{\Omega_{1,N}} x G(x) \;\;\;\;\;\;\;\;\;\;\;\;\;\;\; \alpha > \beta^{\prime}  \; , \; 2 \leq \alpha , \beta \leq N-1 \cr
\displaystyle - \frac{1}{\zeta} q^{t_{2} - t_{1}} \left( \frac{1 - \kappa q\sqrt{\zeta}}{q+1} \right)^2 \Omega_{1,2} \frac{\Omega_{1,N-1}^{2}}{\Omega_{1,N}^{2}} G(x)
\;\;\;\;\;\;\;\;\;\;\;\;\;\;\;\;\;\;\;\;\;\;\;\;\;\;\;\;\;\;\;\;\;\;\; \alpha = 2 \; , \; \beta=1 \cr
\displaystyle  i \frac{q+1}{q-1} \frac{\Omega_{1,m+1} \Omega_{1,m+2}}{\Omega_{1,m+3}} G(x)
\;\;\;\;\;\;\;\;\;\;\;\;\;\;\;\;\;\;\;\;\;\;\;\;\;\;\;\;\;\;\;\;\;\;\;\;\;\;\;\;\;\;\;\;\;\;\;\;\;\;\;\;\;\;\;\;\;\;\; \alpha = 1 \; , \; \beta=m \cr
\displaystyle - \frac{2(q^{2}-1)\sqrt{\zeta}}{(1 + i \kappa (-1)^{m}) (q + i \kappa (-1)^{m} )(1 - \kappa q\sqrt{\zeta})(1 + \kappa \sqrt{\zeta})}\frac{\Omega_{1,2m+2}}{\Omega_{1,m+2}} G(x) \cr
\;\;\;\;\;\;\;\;\;\;\;\;\;\;\;\;\;\;\;\;\;\;\;\;\;\;\;\;\;\;\;\;\;\;\;\;\;\;\;\;\;\;\;\;\;\;\;\;\;\;\;\;\;\;\;\;\;\;\;\;\;\;\;\;\;\;\;\;\;\;\;\;\;\;\;\;\;\;\;\;\;\;\;\;\;\;\;\;\;\;\;\;\; \alpha = 1 \; , \; \beta=m+1 \cr },
\end{equation}
and the parameters $\Omega_{1 , \alpha}$ are required to satisfy the recurrence formula
\begin{equation}
\Omega_{1,\alpha} = -\frac{\Omega_{1,\alpha+1} \Omega_{1,N-\alpha}}{\Omega_{1,N-\alpha+1}}
\;\;\;\;\;\;\;\;\;\;\;\;\;\;\; \alpha = 2, \dots , m-1 .
\end{equation}

Considering now the diagonal entries, they are given by
\begin{equation}
k_{\alpha , \alpha}(x) = \cases{
\displaystyle \left( \frac{2x}{x^{2}-1}-\frac{\Omega_{m+1,m+1} }{x+1} \right) G(x)
\;\;\;\;\;\;\;\;\;\;\;\;\;\;\;\;\;\;\;\;\;\;\;\;\;\;\;\;\;\;\;\;\;\;\;\;\;\; \alpha = 1 \cr
\displaystyle k_{1,1}(x) +  \Omega_{\alpha , \alpha}  G(x)
\;\;\;\;\;\;\;\;\;\;\;\;\;\;\;\;\;\;\;\;\;\;\;\;\;\;\;\;\;\;\;\;\;\;\;\;\;\;\;\;\;\;\;\;\;\;\;\;\;\;\;\;\;\; \alpha = 2, \dots, m+1 \cr
\displaystyle k_{m+1,m+1}(x) \;\;\;\;\;\;\;\;\;\;\;\;\;\;\;\;\;\;\;\;\;\;\;\;\;\;\;\;\;\;\;\;\;\;\;\;\;\;\;\;\;\;\;\;\;\;\;\;\;\;\;\;\;\;\;\;\;\;\;\;\;\;\;\; \alpha = m+2 \cr
\displaystyle k_{m+1,m+1}(x) + \left( \Omega_{\alpha , \alpha} - \Omega_{m+1,m+1} \right) x G(x)
\;\;\;\;\;\;\;\;\;\;\;\;\;\;\;\;\;\;\;\; \alpha = m+3, \dots, N-1 \cr
\displaystyle x^{2}k_{1,1}(x) \;\;\;\;\;\;\;\;\;\;\;\;\;\;\;\;\;\;\;\;\;\;\;\;\;\;\;\;\;\;\;\;\;\;\;\;\;\;\;\;\;\;\;\;\;\;\;\;\;\;\;\;\;\;\;\;\;\;\;\;\;\;\;\;\;\;\;\;\;\;\; \alpha = N \cr}
\end{equation}
where the parameters $\Omega_{\alpha , \alpha}$ are determined by the expressions
\begin{equation}
\Omega_{\alpha ,\alpha} = \cases{
\displaystyle  \Delta_{m} \sum_{k=0}^{\alpha-2}(-\frac{1}{q})^{k} 
\;\;\;\;\;\;\;\;\;\;\;\;\;\;\;\;\;\;\;\;\;\;\;\;\;\;\;\;\;\;\;\;\;\;\;\;\;\;\;\;\;\;\;\;\;\;\;\;\;\;\;\;\;\;\;\;\;\;\;\;\;\;\;\;\;\;\;\;\;\; \alpha = 2 , \dots, m \cr
\displaystyle  \Delta_{m} \left[ \frac{q}{q+1}+(-1)^{m}\frac{1}{q^{m-2}}\frac{q-1}{(q+1)^{2}} \right]
\;\;\;\;\;\;\;\;\;\;\;\;\;\;\;\;\;\;\;\;\;\;\;\;\;\;\;\;\;\;\;\;\;\;\;\; \alpha = m+1 , m + 2 \cr
\displaystyle \Omega_{m+1,m+1} + i \kappa \Delta_{m} \left[ (-1)^{m+1} \frac{q(q-1)}{(q+1)^{2}}+(-1)^{\alpha} \frac{q^{m+4-\alpha}}{q+1} \right] \;\;\;\;\; \alpha = m+3, \dots , N-1 \cr}
\end{equation}
with
\begin{equation}
\Delta_{m} = \frac{2q^{-1}(q+1)^{2}}{(1- i \kappa (-1)^{m})(q + i \kappa (-1)^{m} )(1 + \kappa \sqrt{\zeta})} .
\end{equation}
This solution has altogether $m+1$ free parameters, namely $\Omega_{1 , m+2} , \dots , \Omega_{1, N}$.

\vskip 1cm
\begin{itemize}
\item {\bf Solution $\mathcal{N}_{11}$:}
\end{itemize}
The class of solutions $\mathcal{N}_{11}$ is valid for the vertex model $U_{q}[sl(2n|2)^{(2)}]$ and the corresponding $K$-matrix contains only non-null entries.
The border elements are mainly given by the following expressions
\begin{eqnarray}
k_{\alpha ,N}(x) &=& - \frac{\kappa}{\sqrt{\zeta}} \frac{\epsilon_{\alpha} }{\epsilon_{1}} q^{t_{\alpha}-t_{1}} \Omega_{1,\alpha^{\prime}} x G(x) \;\;\;\;\;\;\;\;\;\;\;\;\;\;\;\;\;\;\;\;\;\;\;\;\;\;\; \alpha = 2, \dots , N-1 \nonumber \\
k_{\alpha ,1}(x) &=& \frac{\epsilon_{\alpha} }{\epsilon_{2}} q^{t_{\alpha}-t_{2}} \frac{\Omega_{2,1} \Omega_{1,\alpha^{\prime}}}{\Omega_{1,N-1}} G(x) \;\;\;\;\;\;\;\;\;\;\;\;\;\;\;\;\;\;\;\;\;\;\;\;\;\;\;\;\;\;\; \alpha = 3 , \dots , N-1 \nonumber \\
k_{N, \alpha}(x) &=& - \frac{\kappa}{\sqrt{\zeta}} \frac{\epsilon_{N} }{\epsilon_{2}}  q^{t_{N}-t_{2}} \frac{\Omega_{2,1} \Omega_{1,\alpha}}{\Omega_{1,N-1}} x G(x) \;\;\;\;\;\;\;\;\;\;\;\;\;\;\;\;\;\;\;\; \alpha = 2, \dots , N-1 \nonumber \\
k_{1, \alpha} (x) &=& \Omega_{1, \alpha} G(x) \;\;\;\;\;\;\;\;\;\;\;\;\;\;\;\;\;\;\;\;\;\;\;\;\;\;\;\;\;\;\;\;\;\;\;\;\;\;\;\;\;\;\;\;\;\;\;\;\;\;\;\;\; \alpha = 2, \dots , N-1
\end{eqnarray}
with $G(x)=x-1$.

The secondary diagonal is constituted by elements $k_{\alpha , \alpha '}(x)$ given by
\begin{equation}
k_{\alpha , \alpha'}(x) = \cases{
\displaystyle \Omega_{1,N} G(x) H_{f}(x) \;\;\;\;\;\;\;\;\;\;\;\;\;\;\;\;\;\;\;\;\;\;\;\;\;\;\;\;\;\;\;\;\;\;\;\;\;\;\;\;\;\;\;\;\;\;\;\;\;\;\;\;\;\;\;\;\;\;\;\;\;\;\;\; \alpha = 1 \cr
\displaystyle \frac{\epsilon_{1} }{\epsilon_{\alpha^{\prime}}} q^{t_{1}-t_{\alpha^{\prime}}} \frac{(1-\kappa q\sqrt{\zeta} )(q+\kappa \sqrt{\zeta}) }{(q^{2}-1)}\frac{\Omega_{1,\alpha^{\prime}}^{2}}{\Omega_{1,N}} G(x) H_{b}(x) 
\;\;\;\;\;\;\;\;\;\;\;\;\;\; \alpha = 2 , \dots , N-1 \cr
\displaystyle \frac{\epsilon_{N-1} }{\epsilon_{2}} q^{t_{N-1}-t_{2}} \frac{\Omega_{1,N} \Omega_{2,1}^{2}}{\Omega_{1,N-1}^{2}} G(x) H_{f}(x)
\;\;\;\;\;\;\;\;\;\;\;\;\;\;\;\;\;\;\;\;\;\;\;\;\;\;\;\;\;\;\;\;\;\;\;\; \alpha =  N \cr },
\end{equation}
where the functions $H_{b}(x)$ and $H_{f}(x)$ were already given in (\ref{hfhb}). The remaining non-diagonal entries are determined by
the expression
\begin{equation}
k_{\alpha , \beta}(x)= \cases{
\displaystyle \frac{\kappa}{\sqrt{\zeta}}  \frac{\epsilon_{\alpha} }{\epsilon_{1}} q^{t_{\alpha} - t_{1}} \left( \frac{1 - \kappa q\sqrt{\zeta} }{q-1} \right) \frac{\Omega_{1,\alpha^{\prime}} \Omega_{1,\beta}}{\Omega_{1,N} } G(x)
\;\;\;\;\;\;\;\;\;\;\;\;\;\;\;\;\;\;\;\; \alpha < \beta^{\prime} \; , \; 2 \leq \alpha ,\beta \leq N-1 \cr
\displaystyle \frac{1}{\zeta} \frac{\epsilon_{\alpha} }{\epsilon_{1}} q^{t_{\alpha} - t_{1} } \left( \frac{1 - \kappa q\sqrt{\zeta}}{q-1} \right) \frac{\Omega_{1,\alpha^{\prime} } \Omega_{1,\beta}}{\Omega_{1,N}} x G(x)
\;\;\;\;\;\;\;\;\;\;\;\;\;\;\;\;\;\;\;\;\; \alpha > \beta^{\prime} \; ,\; 2 \leq \alpha ,\beta \leq N-1 \cr
\displaystyle  \frac{(-1)^{n+1} (q^n + i \kappa)^2 }{\zeta (q^2-1)} \frac{\Omega_{1,n+1} \Omega_{1,n+2} \Omega_{1,N-1}}{\Omega_{1,N}^2} G(x)
\;\;\;\;\;\;\;\;\;\;\;\;\;\;\;\;\;\;\;\;\;\;\;\;\;\;\;\;\;\;\;\;\;\;\; \alpha = 2 \; ,\; \beta=1 \cr
\displaystyle \frac{2 \kappa (q^2-1) (-q)^{n-1}}{(1 + i \kappa (-1)^n ) (q + i \kappa (-1)^n) (q^{n-1} + i \kappa) (q^n + i \kappa)}\frac{\Omega_{1,N} }{\Omega_{1,n+2}} G(x)  \;\;\;\;\;\;\;\;\;\; \alpha = 1 \; ,\; \beta=n+1 \cr},
\end{equation}
and the parameters $\Omega_{1, \alpha}$ are required to satisfy
\begin{equation}
\Omega_{1,\alpha} = - \frac{\Omega_{1,\alpha+1} \Omega_{1,N-\alpha}}{\Omega_{1,N+1-\alpha}}
\;\;\;\;\;\;\;\;\;\;\;\;\;\;\;\;\;\;\;\;\; \alpha = 2 , \dots , n .
\end{equation}

With respect to the diagonal entries, they are given by
\begin{equation}
k_{\alpha , \alpha}(x) = \cases{
\displaystyle \left( \frac{2x}{x^{2}-1}-\frac{\Omega_{n+1,n+1}}{x+1} \right) G(x) 
\;\;\;\;\;\;\;\;\;\;\;\;\;\;\;\;\;\;\;\;\;\;\;\;\;\;\;\;\;\;\;\;\;\;\;\;\; \alpha = 1 \cr
\displaystyle k_{1,1}(x) +  \Omega_{\alpha , \alpha}  G(x)
\;\;\;\;\;\;\;\;\;\;\;\;\;\;\;\;\;\;\;\; \;\;\;\;\;\;\;\;\;\;\;\;\;\;\;\;\;\;\;\;\;\;\;\;\;\;\;\;\;\;\; \alpha = 2, \dots, n+1 \cr
\displaystyle k_{n+1,n+1}(x)
\;\;\;\;\;\;\;\;\;\;\;\;\;\;\;\;\;\;\;\;\;\;\;\;\;\;\;\;\;\;\;\;\;\;\;\;\;\;\;\; \;\;\;\;\;\;\;\;\;\;\;\;\;\;\;\;\;\;\;\;\;\;\;  \alpha = n+2 \cr
\displaystyle k_{n+1,n+1}(x) + \left( \Omega_{\alpha ,\alpha} - \Omega_{n+1,n+1} \right) x G(x)
\;\;\;\;\;\;\;\;\;\;\;\;\;\;\;\;\;\;\;\; \alpha = n+3 , \dots , N-1 \cr
\displaystyle x^{2} k_{1,1}(x) \;\;\;\;\;\;\;\;\;\;\;\;\;\;\;\;\;\;\;\;\;\;\;\;\;\;\;\;\;\;\;\;\;\;\;\;\;\;\;\;\;\;\;\;\;\;\;\;\;\;\;\;\;\;\;\;\;\;\;\;\;\;\;\;\;\;\; \alpha = N \cr}
\end{equation}
and the parameters $\Omega_{\alpha ,\alpha}$ are fixed by the relations
\begin{equation}
\Omega_{\alpha , \alpha}=\cases{
\displaystyle  \frac{ - 2 (-1)^{\alpha + n} \left[   q^{\alpha-1} + q^{\alpha - 2} - (-1)^{\alpha} (q-1) \right] }{(1 - i \kappa (-1)^{n}) (q + i \kappa (-1)^{n}) (i \kappa + q^{n-1})} \;\;\;\;\;\;\;\;\;\;\;\;\;\;\;\;\;\;\;\;\;\;\;\;\;\;\;\;\;\;\;\;\;\;\; \alpha = 2 , \dots , n+1 \cr
\displaystyle \Omega_{n+1,n+1} \;\;\;\;\;\;\;\;\;\;\;\;\;\;\;\;\;\;\;\;\;\;\;\;\;\;\;\;\;\;\;\;\;\;\;\;\;\;\;\;\;\;\;\;\;\;\;\;\;\;\;\;\;\;\;\;\;\;\;\;\;\;\;\;\;\;\;\;\;\;\;\;\;\;\;\;\;\;\;\;\;\;\;\;\;\;\; \alpha = n+2 \cr
\displaystyle \Omega_{n+1,n+1} + \Delta_{n} \sum_{k=0}^{\alpha - n - 3}(-q)^{k}
\;\;\;\;\;\;\;\;\;\;\;\;\;\;\;\;\;\;\;\;\;\;\;\;\;\;\;\;\;\;\;\;\;\;\;\;\;\;\;\;\;\;\;\;\;\;\;\;\;\;\;\;\;\;\;\;\;\;\;  \alpha = n+3 , \dots , N-1 \cr}
\end{equation}
where
\begin{equation}
\Delta_{n} = \frac{2 (-1)^{n+1} (q+1)^2}{(1 + i \kappa (-1)^n )(q + i \kappa (-1)^n )( q^{n-1} + i \kappa)} .
\end{equation}
The variables $\Omega_{1, n+2} , \dots , \Omega_{1,N}$ give us a total amount of $n+1$ free parameters.

\vskip 1cm
\begin{itemize}
\item {\bf Solution $\mathcal{N}_{12}$:}
\end{itemize}

The solution $\mathcal{N}_{12}$ also does not contain null entries and it is valid for the $U_{q}[sl(2n+1|2)^{(2)}]$ vertex model. Considering first the non-diagonal entries, we have the following expression determining border elements,
\begin{eqnarray}
k_{\alpha ,N}(x) &=& - \frac{\kappa}{\sqrt{\zeta}} \frac{\epsilon_{\alpha} }{\epsilon_{1}} q^{t_{\alpha}-t_{1}} \Omega_{1,\alpha^{\prime}} x G(x) \;\;\;\;\;\;\;\;\;\;\;\;\;\;\;\;\;\;\;\;\;\;\;\;\;\;\;\;\;\;\;\;\;\;\; \alpha = 2, \dots , N-1 \nonumber \\
k_{\alpha ,1}(x) &=&  \frac{\epsilon_{\alpha} }{\epsilon_{2}} q^{t_{\alpha}-t_{2}} \frac{\Omega_{2,1} \Omega_{1,\alpha^{\prime}}}{\Omega_{1,N-1}} G(x) \;\;\;\;\;\;\;\;\;\;\;\;\;\;\;\;\;\;\;\;\;\;\;\;\;\;\;\;\;\;\;\;\;\;\;\;\;\;\; \alpha = 3 , \dots , N-1 \nonumber \\
k_{N, \alpha}(x) &=& - \frac{\kappa}{\sqrt{\zeta}} \frac{\epsilon_{N} }{\epsilon_{2}}  q^{t_{N}-t_{2}} \frac{\Omega_{2,1} \Omega_{1,\alpha}}{\Omega_{1,N-1}} x G(x) \;\;\;\;\;\;\;\;\;\;\;\;\;\;\;\;\;\;\;\;\;\;\;\;\;\;\;\;\; \alpha = 2, \dots , N-1 \nonumber \\
k_{1, \alpha} (x) &=& \Omega_{1, \alpha} G(x) \;\;\;\;\;\;\;\;\;\;\;\;\;\;\;\;\;\;\;\;\;\;\;\;\;\;\;\;\;\;\;\;\;\;\;\;\;\;\;\;\;\;\;\;\;\;\;\;\;\;\;\;\;\;\;\;\;\;\;\;\; \alpha = 2, \dots , N-1,
\end{eqnarray}
and the following one for the entries of the secondary diagonal
\begin{equation}
k_{\alpha , \alpha'}(x) =\cases{
\displaystyle \Omega_{1, N} G(x) H_{f}(x) \;\;\;\;\;\;\;\;\;\;\;\;\;\;\;\;\;\;\;\;\;\;\;\;\;\;\;\;\;\;\;\;\;\;\;\;\;\;\;\;\;\;\;\;\;\;\;\;\;\;\;\;\;\;\;\;\;\;\;\;\; \alpha =1 \cr
\displaystyle \frac{\epsilon_{1}}{\epsilon_{\alpha^{\prime}}} q^{t_{1}-t_{\alpha^{\prime}}} \frac{(1-\kappa q\sqrt{\zeta} )(q+\kappa \sqrt{\zeta}) }{(q^{2}-1)}\frac{\Omega_{1,\alpha^{\prime}}^{2}}{\Omega_{1,N}} G(x) H_{b}(x) 
\;\;\;\;\;\;\;\;\;\;\; \alpha = 2 , \dots , N-1 \; , \; \alpha \neq \frac{N-1}{2} \cr
\displaystyle \frac{\epsilon_{N-1}}{\epsilon_{2}} q^{t_{N-1}-t_{2}} \frac{\Omega_{1,N} \Omega_{2,1}^{2}}{\Omega_{1,N-1}^{2}} G(x) H_{f}(x)
\;\;\;\;\;\;\;\;\;\;\;\;\;\;\;\;\;\;\;\;\;\;\;\;\;\;\;\;\;\;\;\;\; \alpha = N, \cr}
\end{equation}
with $H_{b}(x)$ and $H_{f}(x)$ given in (\ref{hfhb}).
The remaining non-diagonal entries are determined by
\begin{equation}
k_{\alpha , \beta}(x) =\cases{
\displaystyle \frac{\kappa}{\sqrt{\zeta}}  \frac{\epsilon_{\alpha} }{\epsilon_{1}} q^{t_{\alpha} - t_{1}} \left( \frac{1 - \kappa q\sqrt{\zeta} }{q-1} \right) \frac{\Omega_{1,\alpha^{\prime}} \Omega_{1,\beta}}{\Omega_{1,N} } G(x)
\;\;\;\;\;\;\;\;\;\;\;\;\;\;\; \alpha < \beta^{\prime} \; , \; 2 \leq \alpha ,\beta \leq N-1 \cr
\displaystyle \frac{1}{\zeta} \frac{\epsilon_{\alpha} }{\epsilon_{1}} q^{t_{\alpha} - t_{1} } \left( \frac{1 - \kappa q\sqrt{\zeta}}{q-1} \right) 
\frac{\Omega_{1,\alpha^{\prime} } \Omega_{1,\beta}}{\Omega_{1,N}} x G(x)
\;\;\;\;\;\;\;\;\;\;\;\;\;\;\;\; \alpha > \beta^{\prime} \; , \; 2 \leq \alpha ,\beta \leq N-1 \cr
\displaystyle \frac{(-1)^n (1 - \kappa q \sqrt{\zeta})^2 }{\zeta (q^2-1) } \frac{\Omega_{1,n+1} \Omega_{1,n+3} \Omega_{1,N-1} }{\Omega_{1,N}^2 } G(x)
\;\;\;\;\;\;\;\;\;\;\;\;\;\;\;\;\;\;\;\;\;\;\;\;\;\;\;\;\;\; \alpha = 2 \; , \; \beta=1 \cr
\displaystyle \left[ \frac{2 \kappa (-1)^n q^{n-\frac{1}{2}} (q-1) (\sqrt{q} - i \kappa (-1)^n) }{(1 - \kappa \sqrt{\zeta}) (1 - \kappa q \sqrt{\zeta} ) (\sqrt{q} + i \kappa (-1)^n) } \frac{\Omega_{1,N} }{ \Omega_{1,n+3}} \right. \cr
\displaystyle \left. \;\;\; - \frac{ i \kappa (-1)^n ( \sqrt{q} - i \kappa (-1)^n ) }{ (\sqrt{q} + i \kappa (-1)^n ) } \frac{\Omega_{1,n+2}^2 }{\Omega_{1,n+3}} \right] G(x)
\;\;\;\;\;\;\;\;\;\;\;\;\;\;\;\;\;\;\;\;\;\;\;\;\;\;\; \alpha = 1 \; , \; \beta=n+1 , \cr}
\end{equation}
and the following recurrence formula should also holds
\begin{equation}
\Omega_{1, \alpha} = - \frac{\Omega_{1,\alpha+1} \Omega_{1,N-\alpha} }{\Omega_{1,N+1-\alpha} }
\;\;\;\;\;\;\;\;\;\;\;\;\;\;\;\;\;\;\;\;\;\;\;\;\;\;\; \alpha = 2 , \dots , n .
\end{equation}

With regard to the diagonal entries, they are given by
\begin{equation}
k_{\alpha ,\alpha}(x) = \cases{
\displaystyle \left( \frac{2 x}{(x^2-1)} - \frac{\Omega_{n+2,n+2} }{(x+1)} \right) G(x)
+ \frac{(1 + \kappa \sqrt{\zeta}) }{(x^2-1)} \Delta(x) 
\;\;\;\;\;\;\;\;\;\;\;\;\;\;\;\;\;\;\;\;\;\; \alpha =1 \cr
\displaystyle k_{1,1}(x) + \Omega_{\alpha , \alpha} G(x)
\;\;\;\;\;\;\;\;\;\;\;\;\;\;\;\;\;\;\;\;\;\;\;\;\;\;\;\;\;\;\;\;\;\;\;\;\;\;\;\;\;\;\;\;\;\;\;\;\;\;\;\;\;\;\;\;\;\;\;\;\;\;\;\;\;\;\;\; \alpha = 2, \dots, n+1 \cr
\displaystyle k_{1,1}(x) + \Omega_{n+2,n+2} G(x) + \Delta(x)
\;\;\;\;\;\;\;\;\;\;\;\;\;\;\;\;\;\;\;\;\;\;\;\;\;\;\;\;\;\;\;\;\;\;\;\;\;\;\;\;\;\;\;\;\;\;\;\;\;\;\;\;\;\;\;\;\;\;\; \alpha = n+2 \cr
\displaystyle k_{n+2,n+2}(x) + \left( \Omega_{n+3,n+3} - \Omega_{n+2,n+2} \right) x G(x) + \kappa \sqrt{\zeta} \Delta(x) 
\;\;\;\;\;\;\;\;\;\;\;\;\;\;\;\;\;\;\;\; \alpha = n+3 \cr
\displaystyle k_{n+3,n+3}(x) + \left( \Omega_{\alpha ,\alpha} - \Omega_{n+3,n+3} \right) x G(x) 
\;\;\;\;\;\;\;\;\;\;\;\;\;\;\;\;\;\;\;\;\;\;\;\;\;\;\;\;\;\; \alpha = n+4 , \dots , N-1 \cr
\displaystyle x^{2} k_{1,1}(x) \;\;\;\;\;\;\;\;\;\;\;\;\;\;\;\;\;\;\;\;\;\;\;\;\;\;\;\;\;\;\;\;\;\;\;\;\;\;\;\;\;\;\;\;\;\;\;\;\;\;\;\;\;\;\;\;\;\;\;\;\;\;\;\;\;\;\;\;\;\;\;\;\;\;\;\;\;\; \alpha = N \cr}
\end{equation}
where 
\begin{equation}
\Delta(x) = i q^{1-n} \frac{(1 - \kappa q \sqrt{\zeta} ) }{(q^2 -1)}
\frac{\Omega_{1, n+1} \Omega_{1, n+3} }{\Omega_{1, N}} (x-1) G(x) .
\end{equation}
In their turn the parameters $\Omega_{\alpha , \alpha}$ are fixed by the following expression
\begin{equation}
\Omega_{\alpha , \alpha} = \cases{
\displaystyle (-1)^{\alpha} Q_{n} \left[ q^{\alpha - 2} (q+1) + (-1)^{\alpha} (1-q) \right] 
\;\;\;\;\;\;\;\;\;\;\;\;\;\;\;\;\;\;\;\;\;\;\;\;\;\;\;\;\;\;\;\;\;\;\;\;\;\; \alpha = 2 , \dots , n+1 \cr
\displaystyle \frac{2}{q+1} + \frac{2 i \kappa \sqrt{q} (-1)^{n} (q-1)^2 \left( q^{n-1} + (-1)^n \right) }{(q^{n - \frac{1}{2}} + i \kappa) ( q^2 - 1 ) (\sqrt{q} + i \kappa (-1)^n)} \cr
\displaystyle \;\; - \frac{i (q^3+1) ( q^{n-\frac{1}{2}} - i \kappa ) (q^{n + \frac{1}{2}} + i \kappa)}{q^{n - \frac{1}{2}} (q^2 - 1) (q^{\frac{3}{2}} - i \kappa (-1)^n ) (\sqrt{q} + i \kappa (-1)^n) } \frac{\Omega_{1,n+2}^2 }{\Omega_{1,N}}
\;\;\;\;\;\;\;\;\;\;\;\;\;\;\;\;\;\;\;\;\;\;\;\;\; \alpha = n+2 \cr
\displaystyle  2 - i \kappa \sqrt{q} Q_{n}  \left[ q^n - q^{n-1} + (-1)^{n+1} (q+1) \right]
\;\;\;\;\;\;\;\;\;\;\;\;\;\;\;\;\;\;\;\;\;\;\;\;\;\;\;\;\;\;\;\;\;\;\;\;\;\;\;\;\;\;\;\; \alpha = n+3 \cr
\displaystyle \Omega_{n+3,n+3} - i \kappa (-1)^n \sqrt{q} (q+1)^2 Q_{n} \sum_{k=0}^{\alpha - n -4} (-q)^k 
\;\;\;\;\;\;\;\;\;\;\;\;\;\;\;\;\;\;\;\; \alpha = n+4 , \dots , N-1 \cr}
\end{equation}
and the auxiliary parameter $Q_{n}$ is given by
\begin{eqnarray}
Q_{n} = - \frac{2}{(1 - \kappa \sqrt{\zeta}) (\sqrt{q} + i \kappa (-1)^n )^2}
+ \frac{i (1 - \kappa q \sqrt{\zeta} ) }{ q^{n - \frac{1}{2}} (q-1) (\sqrt{q} + i \kappa (-1)^n)^2 }\frac{\Omega_{1,n+2}^2 }{ \Omega_{1,N} } .
\end{eqnarray}
This solution possess $n+2$ free parameters, namely $\Omega_{1 , n+2} , \dots , \Omega_{1,N}$.

\vskip 1cm
\section{Concluding Remarks}
In this work we have presented the general set of regular solutions of the graded reflection equation for the
$U_{q}[sl(r|2m)^{(2)}]$ vertex model. Our findings can be summarized into four classes of diagonal solutions
and twelve classes of non-diagonal ones.
These results pave the way to construct, solve and study physical properties of the underlying quantum spin chains
with open boundaries, generalizing the previous efforts made for the case of periodic boundary conditions \cite{GALLEAS1, GALLEAS2}.

Although we expect that the Algebraic Bethe Ansatz solution of the models constructed from the diagonal solutions presented here
can be obtained by adapting the results of \cite{G4}, the algebraic-functional method presented in \cite{GALLEAS4} may be
a possibility to treat the non-diagonal cases.

For further research, an interesting possibility would be the investigation of soliton non-preserving boundary conditions \cite{G3,DO1} for quantum spin chains based on $q$-deformed Lie algebras and superalgebras, which could also be performed by adapting the method described in \cite{LIM}. We expect the results presented here to motivate further developments on the subject of integrable open boundaries for vertex models based on $q$-deformed Lie superalgebras. In particular, the classification of the solutions of the graded reflection equation for 
others $q$-deformed Lie superalgebras, which we hope to report in a future work.

\vskip 2cm
\section{Acknowledgments} 
The work of A. Lima-Santos is partially supported by the Brazilian research councils CNPq and FAPESP.
W. Galleas thanks the agency FAPESP for the financial support.

\newpage
\section*{\bf Appendix A: The $U_{q}[sl(1|2)^{(2)}]$ case}
\setcounter{equation}{0}
\renewcommand{\theequation}{A.\arabic{equation}}

The reflection equation associated with the $U_{q}[sl(1|2)^{(2)}]$ vertex model admits more general solutions than the
corresponding ones obtained from the general series presented in the section 3. In this case the reflection matrices were previously
studied in \cite{LIM1} and we have obtained the following solutions
\begin{equation}
\label{a12}
K^{-}(x)=\pmatrix{
\frac{\Omega (x^{-1}-1)+2}{\Omega (x-1)+2} & 0 & 0 \cr
0 & 1 & 0 \cr
0 & 0 & \frac{\Omega x(xq+1)-2x}{\Omega (x+q)-2x} \cr}
\end{equation}
\begin{equation}
\label{b12}
K^{-}(x)=\pmatrix{
1 & 0 & \frac{\Omega (x^{2}-1)}{2} \cr 
0 & \frac{x^{2}q+1}{q+1} & 0 \cr
-\frac{2q(x^{2}-1)}{\Omega (q+1)^{2}} & 0 & x^{2} \cr}
\end{equation}
\n where $\Omega$ is a free parameter.

In addition to the solutions (\ref{a12}) and (\ref{b12}) we also have a solution in the general form
\begin{equation}
K^{-}(x)=\pmatrix{
k_{1,1}(x) & k_{1,2}(x) & k_{1,3}(x) \cr
k_{2,1}(x) & k_{2,2}(x) & k_{2,3}(x) \cr
k_{3,1}(x) & k_{3,2}(x) & k_{3,3}(x) \cr}.
\end{equation}
The diagonal entries are then given by
\begin{eqnarray}
k_{1,1}(x) &=&\frac{2(x+q)(x-1)}{(x^{2}+q)(x^{2}-1)}
-(\Omega_{1,2}+i\sqrt{q}\Omega_{2,3})\frac{\left[q(x-1)\Omega_{2,3}-i\sqrt{q}
(x+q)\Omega_{1,2}\right]}{(q-1)(x+1)(q+x^{2})}\frac{(x-1)}{\Omega_{1,3}} \nonumber \\
&-&\frac{\left[(q+1)(\sqrt{q}\Omega_{2,3}+ix\Omega_{1,2})\Omega_{1,3}\Omega_{2,1}-2iq\Omega_{1,2}\Omega_{2,3}\right]}{\sqrt{q}(x+1)(q+x^{2})}\frac{(x-1)}{\Omega_{1,2}^{2}} \nonumber \\
k_{2,2}(x) &=&-xk_{1,1}(x)+\frac{2x}{(q+x^{2})}\left[ q+x+i\sqrt{q}(x-1)
\frac{\Omega_{2,3}}{\Omega_{1,2}}\right]  \nonumber \\
&-&\left[ i\sqrt{q}(x^{2}-1)+x(q+1)\frac{\Omega_{2,3}}{\Omega_{1,2}}\right] 
\frac{\Omega_{2,1}}{\Omega_{1,2}}\frac{\Omega_{1,3} (x-1)}{(q+x^{2})} \nonumber \\
k_{3,3}(x) &=& x^{2}k_{1,1}(x)+\left[ \Omega_{3,3}-2+i\frac{\sqrt{q}
(x-1)(x-q)}{(q-1)(q+x^{2})}\frac{\Omega_{1,2}^{2}-\Omega_{2,3}^{2}}{\Omega_{1,3}
}\right] x (x-1) ,
\end{eqnarray}
and the remaining elements can be written as
\begin{eqnarray}
k_{1,2}(x) &=&\frac{(q+x)\Omega_{1,2}+i(x-1)\sqrt{q}\Omega_{2,3}}{q+x^{2}} (x-1)  \;\;\;\;\;\;\;\;\;\;
k_{1,3}(x) = \Omega_{1,3} (x-1) \nonumber \\
k_{2,1}(x) &=&\frac{(q+x)\Omega_{2,1}+i(x-1)\sqrt{q}\Omega_{3,2}}{q+x^{2}} (x-1)
\;\;\;\;\;\;\;\;\;\; 
k_{2,3}(x) = \frac{(q+x)\Omega_{2,3}+i(x-1)\sqrt{q}\Omega_{1,2}}{q+x^{2}} x (x-1) \nonumber \\
k_{3,2}(x) &=&\frac{(q+x)\Omega_{3,2}+i(x-1)\sqrt{q}\Omega_{2,1}}{q+x^{2}} x (x-1)
\;\;\;\;\;\;\;\; 
k_{3,1}(x) = \Omega_{3,1} (x-1) .
\end{eqnarray}
This solution has altogether three free parameters $\Omega_{1,2}, \Omega_{1,3}$ and $\Omega_{2,1}$ and the remaining variables $\Omega_{\alpha , \beta}$ are determined by
\begin{eqnarray}
\Omega_{2,3} &=&-2\frac{q-1}{q+1}\frac{\Omega_{1,3}}{\Omega_{1,2}}-i\frac{q-1}{
\sqrt{q}}\Omega_{2,1}\frac{\Omega_{1,3}^{2}}{\Omega_{1,2}^{2}}
\;\;\;\;\;\;\;\;\;\;\;\;
\Omega_{3,2} = \Omega_{2,3}\frac{\Omega_{2,1}}{\Omega_{1,2}} \nonumber \\
\Omega_{3,3} &=& 2+i\frac{\sqrt{q}}{q-1}\frac{\Omega_{2,3}^{2}-\Omega_{1,2}^{2}}{\Omega_{1,3}} 
\;\;\;\;\;\;\;\;\;\;\;\;\;\;\;\;\;\;\;\;\;\;\;\;\;\;
\Omega_{3,1}=\Omega_{1,3}\frac{\Omega_{2,1}^{2}}{\Omega_{1,2}^{2}}.
\end{eqnarray}

\section*{\bf Appendix B: The $U_{q}[sl(2|2)^{(2)}]$ case}
\setcounter{equation}{0}
\renewcommand{\theequation}{B.\arabic{equation}}
The set of $K$-matrices associated with the $U_{q}[sl(2|2)^{(2)}]$ vertex model includes 
both diagonal and non-diagonal solutions. The solutions intrinsically diagonal contain only one free parameter $\Omega$ and they
are given by
\begin{equation}
K^{-}(x)=\pmatrix{
x^{-1} &  &  &  \cr
& -\frac{\Omega (x-1)+2}{\Omega (x-1)-2x} &  &  \cr
&  & \frac{\Omega (x+q^{2})-2x}{\Omega (xq^{2}+1)-2} &  \cr
&  &  & x \cr}
\end{equation}
and
\begin{equation}
K^{-}(x)=\pmatrix{
x^{-1} &  &  &  \cr 
& -\frac{\Omega (x-1)+2}{\Omega (x-1)-2x} &  &  \cr 
&  & -\frac{\Omega (x-1)+2}{\Omega (x-1)-2x} &  \cr
&  &  & -\left[\frac{\Omega (x-1)+2}{\Omega (x-1)-2x} \right] \left[\frac{\Omega (xq^{2}+1)-2}{\Omega
(x+q^{2})-2x}\right]x \cr}.
\end{equation}
We have also found the following non-diagonal solutions
\begin{eqnarray}
K^{-}(x)&=&\pmatrix{ 
1 & 0 & 0 & \frac{\Omega_{2}}{2}(x^{2}-1) \cr
0 & \frac{\Omega_{1} }{2}(x^{2}-1)+1 & 0 & 0 \cr
0 & 0 & -\frac{\Omega_{1} }{2}(x^{2}-1)+x^{2} & 0 \cr
\frac{\Omega_{1} (\Omega_{1} -2)}{2\Omega_{2}}(x^{2}-1) & 0 & 0 & x^{2} \cr} \\
\nonumber \\
K^{-}(x)&=&\pmatrix{ 
-\frac{\Omega_{1} (x^{2}-1)-2x^{2}}{2x} & 0 & 0 & 0 \cr 
0 & x & \frac{\Omega_{2}}{2}(x^{2}-1) & 0 \cr
0 & \frac{\Omega_{1} (\Omega_{1} -2)}{2\Omega_{2}}(x^{2}-1) & x & 0 \cr 
0 & 0 & 0 & \frac{x}{2}[\Omega_{1} (x^{2}-1)+2] \cr}
\end{eqnarray}
containing two free parameters $\Omega_{1}$ and $\Omega_{2}$, and one solution in the form
\begin{equation}
\label{b5}
K^{-}(x)=\pmatrix{
k_{1,1}(x) & k_{1,2}(x) & k_{1,3}(x) & k_{1,4}(x) \cr 
k_{2,1}(x) & k_{2,2}(x) & k_{2,3}(x) & k_{2,4}(x) \cr
k_{3,1}(x) & k_{3,2}(x) & k_{3,3}(x) & k_{3,4}(x) \cr
k_{4,1}(x) & k_{4,2}(x) & k_{4,3}(x) & k_{4,4}(x) \cr}.
\end{equation}
Concerning the solution (\ref{b5}), the diagonal entries are given by
\begin{eqnarray}
k_{1,1}(x) &=&\left( \frac{xq-i\kappa }{q-i\kappa }\right) \;\;\;\;\;\;\;\;\;\;\;\; k_{2,2}(x)=x\left( \frac{x+i\kappa q}{1+i\kappa q}\right) \nonumber \\
k_{3,3}(x) &=&x\left( \frac{x+i\kappa q}{1+i\kappa q}\right) \;\;\;\;\;\;\;\;\; k_{4,4}(x)=x^{2}\left( \frac{xq-i\kappa }{q-i\kappa}\right) \end{eqnarray}
while the non-diagonal entries are given by the following expressions,
\begin{eqnarray}
k_{1,2}(x) &=&\kappa \frac{\Omega_{2}}{2\Omega_{1}}\frac{q^{2}-1}{q^{2}+1}(x^{2}-1) \;\;\;\;\;\;\;\;\;\;\;\;\;\;\;\;\;\;\;\;\;\;\;\;\;\;\;\;\;\;\;\;
k_{2,1}(x) =-\kappa \frac{\Omega_{1}}{2\Omega_{2}}\frac{q+i\kappa }{q-i\kappa }(x^{2}-1) \nonumber \\
k_{1,3}(x) &=& \frac{\Omega_{1}}{2}(x^{2}-1) \;\;\;\;\;\;\;\;\;\;\;\;\;\;\;\;\;\;\;\;\;\;\;\;\;\;\;\;\;\;\;\;\;\;\;\;\;\;\;\;\;\;\;\;\;\;\;
k_{2,3}(x) =-\kappa \frac{\Omega_{1}^{2}}{2\Omega_{2}}\frac{(qx+i\kappa )(q+i\kappa )}{q^{2}-1}(x^{2}-1) \nonumber \\
k_{1,4}(x) &=& \frac{\Omega_{2}}{2}\frac{x-i\kappa q}{1-i\kappa q}(x^{2}-1) \;\;\;\;\;\;\;\;\;\;\;\;\;\;\;\;\;\;\;\;\;\;\;\;\;\;\;\;\;\;\;\;\;\;\;
k_{2,4}(x) =-\kappa \frac{\Omega_{1}}{2}x(x^{2}-1) \nonumber \\
k_{3,1}(x) &=&-\frac{1}{2\Omega_{1}}\frac{q^{2}-1}{(q-i\kappa )^{2}}(x^{2}-1) \;\;\;\;\;\;\;\;\;\;\;\;\;\;\;\;\;\;\;\;\;\;\;\;\;\;\;\;
k_{4,1}(x) =\frac{1}{2\Omega_{2}}\frac{x-i\kappa q}{1-i\kappa q}\left( \frac{q+i\kappa }{q-i\kappa }\right) ^{2}(x^{2}-1) \nonumber \\
k_{3,2}(x) &=&-\kappa \frac{\Omega_{2}}{2\Omega_{1}^{2}}\frac{qx+i\kappa }{q+i\kappa }\frac{q^{2}-1}{(q-i\kappa )^{2}}(x^{2}-1) 
\;\;\;\;\;\;\;\;\;\;\;\;\;\;
k_{4,2}(x) =\kappa \frac{1}{2\Omega_{1}}\frac{q^{2}-1}{(q-i\kappa )^{2}}x(x^{2}-1) \nonumber \\
k_{3,4}(x) &=& -\frac{\Omega_{2}}{2\Omega_{1}}\frac{q^{2}-1}{q^{2}+1}x(x^{2}-1) \;\;\;\;\;\;\;\;\;\;\;\;\;\;\;\;\;\;\;\;\;\;\;\;\;\;\;\;\;\;\;
k_{4,3}(x) = \frac{\Omega_{1}}{2\Omega_{2}}\frac{q+i\kappa }{q-i\kappa }x(x^{2}-1) ,
\end{eqnarray}
where $\kappa = \pm 1$ and $\Omega_{1}$ and $\Omega_{2}$ are free parameters.


\begin{thebibliography}{}
\bibitem{POL} A.M. Polyakov, {\it JETP Lett. 12, (1970) 381.}
\bibitem{BEL} A.A. Belavin, A.M. Polyakov and A.B. Zamolodchikov, {\it Nucl. Phys. B 241, (1984) 333.}
\bibitem{CAR} J.L. Cardy, {\it Phase transitions and critical phenomena}, vol 11, Ed. C. Domb and J.L. Lebowitz, New York Academic, (1987).
\bibitem{ADS} J.M. Maldacena, {\it Adv. Theor. Math. Phys. 2, (1998) 231.}
\bibitem{MIN} J.A. Minahan and K. Zarembo, {\it JHEP 0303, (2003) 013.}
\bibitem{BEI} N. Beisert and M. Staudacher, {\it Nucl. Phys. B670, (2003) 439.}
\bibitem{BPR} I. Bena, J. Polchinsky and R. Roiban, {\it Phys. Rev. D69, (2004) 046002.}
\bibitem{JJ} G. Arutyunov, S. Frolov, J. Russo and A.A. Tseytlin, {\it Nucl. Phys. B671, (2003) 3};
G. Arutyunov, J. Russo and A.A. Tseytlin, {\it Phys. Rev. D69, (2004) 086009};
G. Arutyunov and M. Staudacher, {\it JHEP 0403, (2004) 004};
N. Berkovits, {\it JHEP 0503, (2005) 041.}
\bibitem{BAX} R.J. Baxter, {\it Exactly solved models in statistical mechanics}, Academic Press, New York, (1982).
\bibitem{QISM} L.D. Takhtajan and L.D. Faddeev, {\it Russian Math. Surveys 34, (1979) 11.}
\bibitem{KOR} V.E. Korepin, G. Izergin and N.M. Bogoliubov, {\it Quantum Inverse Scattering Method and
Correlation Functions}, Cambridge Univ. Press, Cambridge, (1993).
\bibitem{CAR1} J.L. Cardy, {\it Nucl. Phys. B 275, (1986) 200.}
\bibitem{WIT} E. Witten, {\it JHEP 9807, (1998) 006}; 
J. McGreevy, L. Susskind and N. Tombas, {\it JHEP 0006, (2000) 008.}
\bibitem{CHEN} B. Chen, X.J. Wang and Y.S. Wu, {\it Phys. Lett. B591, (2004) 170.}
\bibitem{DEW} O. DeWolfe and N. Mann, {\it JHEP 0404, (2004) 035.}
\bibitem{BER} D. Berenstein and S.E. Vazquez, {\it JHEP 0506, (2005) 059.}
\bibitem{SK} E.K. Sklyanin, {\it J. Phys. A: Math. Gen. 21, (1988) 2375.}
\bibitem{CHER} I.V. Cherednik, {\it Theor. Math. Phys. 61, (1984) 977.}
\bibitem{BAZ} V.V. Bazhanov, {\it Phys. Lett. B 159, (1985) 321.}
\bibitem{JIM} M. Jimbo, {\it Comm. Math. Phys. 102, (1986) 247.}
\bibitem{LIM} R. Malara and A. Lima-Santos, {\it J. Stat. Mech.: Theor. Exp., (2006) P09013.}
\bibitem{G1} G.L. Li, R.H. Yue and B.Y. Hou, {\it Nucl. Phys. B 586, (2000) 711};
A. Gonzalez-Ruiz, {\it Nucl. Phys. B 424, (1994) 468}.
\bibitem{BRA} A.J. Bracken, X.Y. Ge, Y.Z. Zhang and H.Q. Zhou, {\it Nucl. Phys. B 516, (1998) 588}.
\bibitem{GUA} M.J. Martins and X.W. Guan, {\it Nucl. Phys. B 562, (1999) 433.}
\bibitem{G2} D. Arnaudon, J. Avan, N. Crampe, A. Doikou, L. Frappat and E. Ragoucy, {\it 
Nucl. Phys. B 668, (2003) 469};
G.L. Li, K.J. Shi and R.H. Yue, {\it Nucl. Phys. B 687, (2004) 220.}
\bibitem{G3} D. Arnaudon, J. Avan, N. Crampe, A. Doikou, L. Frappat and E. Ragoucy, {\it J. Stat. Mech.: Theor. Exp., (2004) P08005.}
\bibitem{GALLEAS5} W. Galleas, {\it Nucl. Phys. B 777, (2007) 352.}
\bibitem{SHA} V.V. Bazhanov and A.G. Shadrikov, {\it Theor. Math. Phys. 73, (1987) 1302.}
\bibitem{GALLEAS1} W. Galleas and M.J. Martins, {\it Nucl. Phys. B 699, (2004) 455.}
\bibitem{GALLEAS2} W. Galleas and M.J. Martins, {\it Nucl. Phys. B 732, (2006) 444.}
\bibitem{GALLEAS3} W. Galleas and M.J. Martins, {\it Nucl. Phys. B 768, (2007) 219.}
\bibitem{MEZ} L. Mezincescu and R.I. Nepomechie, {\it J. Phys. A: Math. Gen. 24, (1991) 17}; 
L. Mezincescu and R.I. Nepomechie, {\it Int. J. Mod. Phys. A6, (1991) 5231}.
\bibitem{G4} G.L. Li and K.J. Shi, {\it J. Stat. Mech.: Theor. Exp., (2007) P01018.}
\bibitem{GALLEAS4} W. Galleas, {\it Nucl. Phys. B 790, (2008) 524.}
\bibitem{DO1} A. Doikou, {\it J. Phys. A: Math. Gen. 33, (2000) 8797.}
\bibitem{LIM1} A. Lima-Santos, {\it Nucl. Phys. B 558, (1999) 637.}
\end{thebibliography}
\end{document}